\documentclass[twocolumn,showpacs,preprintnumbers,amsmath,amssymb]{revtex4-1}

\usepackage{color}
\usepackage{graphicx}
\usepackage{dcolumn}
\usepackage{bm}

\def\colourcyan#1{\color{black}{#1}}

\begin{document}

\preprint{}

\title{Linear programming analysis of the $R$-parity violation within EDM-constraints}

\author{Nodoka Yamanaka}
  \email{nodoka.yamanaka@riken.jp}
  \affiliation{iTHES Research Group, RIKEN, Wako, Saitama 351-0198, Japan }

\author{Toru Sato}
\affiliation{%
Department of Physics, Osaka University, Toyonaka, Osaka 560-0043, Japan}%

\author{Takahiro Kubota}
\affiliation{%
CELAS, Osaka University, Toyonaka, Osaka 560-0043, Japan, 
Department of Physics, Osaka University, Toyonaka, Osaka 560-0043, Japan, 
\\
Kavli IPMU (WPI), the University of Tokyo, 5-1-5 Kashiwa-no-Ha, Kashiwa City, Chiba 277-8583, Japan
}%

\date{\today}

\begin{abstract}
The constraint on the $R$-parity violating supersymmetric interactions is discussed  in the light of current experimental data of the  electric dipole moment of neutron, $^{129}$Xe , $^{205}$Tl, and  $^{199}$Hg atoms, and YbF and ThO molecules.
To investigate the constraints without relying upon the assumption of the dominance of a particular combination of couplings over all the rest, an extensive use is made of the linear programming method in the scan of the parameter space.
We give maximally possible values for the EDMs of the proton, deuteron, $^3$He nucleus, 
$^{211}$Rn, $^{225}$Ra, $^{210}$Fr, and the $R$-correlation of the neutron beta decay within the constraints from the current experimental data of the EDMs of neutron, $^{129}$Xe, $^{205}$Tl, and  $^{199}$Hg atoms, and YbF and ThO molecules
using the linear programming method.
It is found that the $R$-correlation of the neutron beta decay and hadronic EDMs are very useful observables to constrain definite regions of the parameter space of the $R$-parity violating supersymmetry.
\end{abstract}

\pacs{12.60.Jv, 11.30.Er, 13.40.Em, 14.80.Ly}

\maketitle

\section{\label{sec:intro}Introduction}

The supersymmetric (SUSY) extension of the Standard 
Model (SM) has widely been discussed as a good candidate of the new physics \cite{mssm}.
One of the important aspects of the supersymmetric SM is that it allows room for baryon and lepton number violation. 
The conservation of $R$-{\it parity}  is often introduced to forbid such 
 violation. We must say, however,  that this conservation 
has never been put on a strong convincing basis, and many phenomenological analyses of the $R$-parity violation were done so far 
\cite{rpvphenomenology}.

In spite of dedicated efforts in LHC experiment, an evidence 
 of suparparticles is yet to come. The LHC data have so far placed only  
tight constraints on the SUSY parameter space.
It is to be noted, however,  
 that most of the SUSY analyses of LHC results have been 
performed, assuming the $R$-parity conservation. If this assumption is relaxed, 
decay modes of superparticles become different and the constrained parameter 
space could be significantly altered.
The principal  reason for our renewed interest in RPV SUSY models is to 
extend the scope  of looking at LHC  data  and of groping our
way towards new physics.

One of the promising experimental approach to search for new physics beyond the SM is the electric dipole moment (EDM) \cite{edmreview,khriplovichbook,ginges,pospelovreview,yamanakabook}.
The advantages of the EDM are almost obvious and far-reaching.
Namely, the EDM can be measured with high accuracy in a variety of systems. 
The representative experimental data are the EDMs of $^{129}$Xe atom ($d_{\rm Xe} < 4.0 \times 10^{-27} e\, {\rm cm}$) \cite{rosenberry}, $^{205}$Tl atom ($d_{\rm Tl} < 9 \times 10^{-25} e\, {\rm cm}$) \cite{regan}, neutron ($d_n < 2.9 \times 10^{-26} e\, {\rm cm}$) \cite{baker}, $^{199}$Hg atom ($d_{\rm Tl} < 3.1 \times 10^{-29} e\, {\rm cm}$) \cite{griffith}, muon ($d_\mu < 1.8 \times 10^{-19} e\, {\rm cm}$) \cite{muong2}, YbF molecule ($d_e < 1.05 \times 10^{-27} e\, {\rm cm}$) \cite{hudson}, ThO molecule ($d_e < 8.7 \times 10^{-29} e\, {\rm cm}$) \cite{acme}.
There are also many future experimental prospects such as the measurement of the EDMs of proton \cite{storage,bnl}, deuteron \cite{storage,bnl}, muon \cite{storage,bnl}, $^{225}$Ra atom \cite{mueller}, neutron \cite{ucn}, $^{129}$Xe atom \cite{xeasahi}, $^{210}$Fr atom \cite{sakemi}, etc.
All these EDMs are expected to receive very small contributions from SM \cite{smedm}, which makes them to be an excellent probe of the new physics.
The EDM is so sensitive to the new physics that the supersymmetric models with \cite{susyedm1-loop,susyedm2-loop, susyedmgeneral,falk,hisano,degrassi,susyedmflavorchange,yamanakarainbow1,pospelovreview,ellisgeometricapproach,dekens}
 and without $R$-parity \cite{barbieri,godbole,chang,herczege-n,choi,cch,faessler,yamanaka1,rpvedmsfermion} have been analyzed to a considerable extent, and many supersymmetric CP phases were constrained so far.

The parameter space of the $R$-parity violation is quite large, and the analysis of the whole parameter space including the usual $R$-parity conserving parameters is discouragingly difficult.
In such situations, we often restrict the parameter space only to few parameters to allow for feasible phenomenological analyses, as done in many previous works 
(under the assumptions of a single coupling dominance), and many tight constraints on the RPV couplings have been derived so far.
This approach assuming a single coupling dominance, however, cannot exhaust all corners of the RPV parameter space in which interferences could occur, and thereby some couplings may be sufficiently larger than upper limits derived with this assumption.
In the $R$-parity conserving sector, a systematic analysis of the SUSY CP phases was done by Ellis {\it et al.} 
(for the minimally flavor violating maximally CP violating model), and they obtained a possible large prediction for many prepared experiments such as the EDMs of the deuteron or $^{225}$Ra atom \cite{ellisgeometricapproach}.
If the supersymmetric theory is extended with RPV, an equally full analysis for RPV interactions seems to be needed.
Our aim is then to do a systematic analysis of the full space of the CP violating RPV interactions on the basis of the {\it linear programming method} by using the constraints due to the existing experimental data 
(neutron, $^{129}$Xe, $^{205}$Tl,  $^{199}$Hg, YbF and ThO EDMs plus other CP conserving experimental data of fundamental precision tests \cite{rpvphenomenology}), and to present the maximal expectations for P, CP-odd observables in preparation.
In the present analysis, we will predict the EDMs of the proton, deuteron, $^3$He nucleus, $^{210}$Fr, $^{211}$Rn, $^{225}$Ra atoms, muon, and the $R$-correlation of the neutron beta decay with the linear programming method.

This paper is organized as follows.
We first present in Section \ref{sec:RPV} the RPV interactions and their contribution to the EDM observables.
The elementary level RPV processes as well as the hadronic and many-body physics needed in the computation are explained in detail.
We then give a brief review of the linear programming algorithm in Section \ref{sec:linearprogramming}.
In Section \ref{sec:setup}, we provide the whole setup of the input parameters.
There the complete formulae of the EDMs used in this analysis are shown.
We then analyze the constraints to RPV couplings in Section \ref{sec:analysis}, and also the prospective values of the future EDM experiments.
The last part is devoted to the summary.

\section{\label{sec:RPV}RPV contributions to the EDMs}

\subsection{\label{sec:elementary}Elementary level contribution}

We first review the RPV contribution at the elementary level.
The derivation of the RPV contribution to the EDM observables is based on our previous papers \cite{yamanaka1,rpvedmsfermion}.

The RPV superpotential relevant in our discussion is the following:
\begin{equation}
W_{{\rm R}\hspace{-.5em}/} =
\frac{1}{2} \lambda_{ijk} \epsilon_{ab} L_i^a L_j^b (E^c)_k
+\lambda'_{ijk} \epsilon_{ab} L_i^a Q_j^b ( D^c)_k \ ,
\label{eq:superpotential}
\end{equation}
with $i,j,k=1,2,3$ indicating the generation, $a,b=1,2$ the $SU(2)_L$ indices. 
$L$ and $E^c$ denote the lepton doublet and singlet left-chiral superfields, respectively. 
$Q$ and $D^c$ denote respectively the quark doublet and down quark singlet left-chiral superfields.

In the presence of the bilinear RPV interactions, the authors of \cite{choi} made a suitable use of the flavor basis in which only one of the four $Y=-1/2$ doublet fields bears vacuum expectation value \cite{SVP}.
In this work, however, we assume that the bilinear RPV interactions are absent.
There could also be baryon number violating RPV interactions, but they were omitted in Eq. (\ref{eq:superpotential}) to avoid rapid proton decay.

In connection with the choice of flavor basis, we also note that 
the RPV interactions (\ref{eq:superpotential})  give rise to a new aspect 
in the neutrino mass matrix.  
Majorana neutrino masses are generated by loop diagrams 
due to  (\ref{eq:superpotential}), 
in which  $d$-quark and $d$-squark are encircling. The coupling constants 
$\lambda _{ijk}^{'}$ are responsible for these mass terms and it has been 
argued in \cite{sallydawson}
 that $\lambda ^{'}_{133}$
is strongly constrained. In principle we are always using the mass basis 
for quarks and leptons in  (\ref{eq:superpotential}),  and the mixing matrices 
necessarily show up when we evaluate loop diagrams. In practice, however, 
our numerical calculations are insensitive to  the neutrino mixing matrix,  
since we will always assume common values for the squark and 
slepton masses.

\begin{figure}[htb]
\includegraphics[width=8cm]{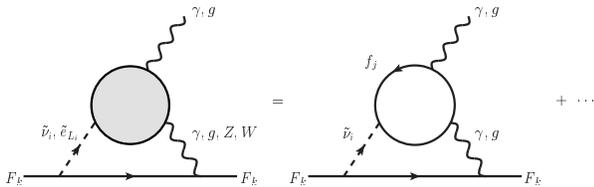}
\caption{\label{fig:barr-zee} 
Barr-Zee type two-loop level contribution to the fermion EDM.
}
\end{figure}

The RPV lagrangian that follows from the superpotential (\ref{eq:superpotential}) is then
\begin{eqnarray}
{\cal L }_{\rm R\hspace{-.5em}/\,} &=&
- \frac{1}{2} \lambda_{ijk} \Bigl[
\tilde \nu_i \bar e_k P_L e_j +\tilde e_{Lj} \bar e_k P_L \nu_i + \tilde e_{Rk}^\dagger \bar \nu_i^c P_L e_j 
\nonumber\\
&& \hspace{17em}
-(i \leftrightarrow j ) \Bigr] 
\nonumber\\
&&-\lambda'_{ijk} \Bigl[
\tilde \nu_i \bar d_k P_L d_j + \tilde d_{Lj} \bar d_k P_L \nu_i +\tilde d_{Rk}^\dagger \bar \nu_i^c P_L d_j 
\nonumber\\
&& \hspace{4em}
-\tilde e_{Li} \bar d_k P_L u_j - \tilde u_{Lj} \bar d_k P_L e_i - \tilde d_{Rk}^\dagger \bar e_i^c P_L u_j \Bigr] 
\nonumber\\
&&
- \frac{1}{2} \lambda_{ijk} \Bigl[ 
( m_{e_j} \tilde \nu_i \, \tilde e_{Rj} - m_{e_i} \tilde \nu_j \, \tilde e_{Ri})\, \tilde e_{Rk}^\dagger
\nonumber\\
&& \hspace{5em}
+m_{e_k} (\tilde \nu_i \, \tilde e_{Lj} - \tilde \nu_j \, \tilde e_{Li})\, \tilde e_{Lk}^\dagger
\Bigr] 
\nonumber\\
&&
- \lambda'_{ijk} \Bigl[ 
m_{d_j} \tilde \nu_i \, \tilde d_{Rk}^\dagger \tilde d_{Rj}
+m_{d_k} (\tilde \nu_i \, \tilde d_{Lj} - \tilde e_{Li}\, \tilde u_{Lj} )\, \tilde d_{Lk}^\dagger
\nonumber\\
&& \hspace{6.5em}
-m_{u_j} \tilde e_{Li} \tilde d_{Rk}^\dagger \tilde u_{Rj}
-m_{e_i} \tilde u_{Lj} \tilde d_{Rk}^\dagger \tilde e_{Ri}
\Bigr] 
\nonumber\\
&&
+ ({\rm h.c.}) \ .
\label{eq:rpvlagrangian}
\end{eqnarray}
These RPV interactions are of Yukawa type, and are lepton number violating.
The RPV interactions with CP phase contribute to those of the fermion EDM 
\begin{equation}
{\cal L}_{\rm EDM} 
=
 - \frac{i}{2} d_F \bar \psi_F \gamma_5 \sigma^{\mu \nu} \psi_F F_{\mu \nu} 
\, , 
\label{eq:ferminonEDM}
\end{equation}
and the quark chromo-EDM
\begin{equation}
{\cal L}_{\rm cEDM} 
=
 - \frac{i}{2} d^c_q g_s \bar \psi_q \gamma_5 \sigma^{\mu \nu} t_a \psi_q G_{\mu \nu}^a 
\, , 
\label{eq:cedm}
\end{equation}
from the two-loop level Barr-Zee type diagrams  (see Fig. \ref{fig:barr-zee}) \cite{susyedm2-loop,barr-zee,godbole,chang,faessler,yamanaka1,rpvedmsfermion}.
Here the coefficients in (\ref{eq:ferminonEDM}) and (\ref{eq:cedm}) are expressed in terms of the RPV couplings as
\begin{eqnarray}
d_F
&=&
{\rm Im} (\hat \lambda_{ijj} \tilde \lambda^*_{ikk}) d_{\rm BZ} (j,k,Q_f , Q_F)
,
\label{eq:bzfunction}
\\
d^c_q
&=&
{\rm Im} (\lambda'_{ijj} \lambda'^*_{ikk}) d^c_{\rm BZ} (j,k)
,
\label{eq:bzfunction2}
\end{eqnarray}
where $d_{\rm BZ}$ and $d^c_{\rm BZ}$ are defined in Eqs. (\ref{eq:Barr-Zee1}) and  (\ref{eq:chromo-Barr-Zee1}) of Appendix \ref{sec:barr-zee}, respectively.
The indices $j$ and $k$ are the generation of the inner loop and external fermions, respectively.
The RPV couplings $\hat \lambda_{ijj}$ and $\tilde \lambda_{ikk}$ are defined as follows:
$\hat \lambda_{ijj} = \lambda_{ijj}$ if the inner loop fermion is a lepton, $\hat \lambda_{ijj} = \lambda'_{ijj}$ if the inner loop fermion is a (down-type) quark,
$\tilde \lambda^*_{ikk}=\lambda^*_{ikk}$ if $F$ is a lepton, and $\tilde \lambda^*_{ikk}=\lambda'^*_{ikk}$ if $F$ is a (down-type) quark.
It should be noted that, if there were the bilinear RPV interactions, the leading RPV effect would appear at the one-loop level \cite{choi,cch}.
\begin{figure}[htb]
\includegraphics[width=4cm]{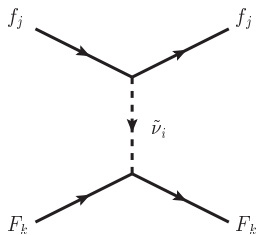}
\caption{\label{fig:4-fermion} 
RPV contribution to the P, CP-odd four-fermion interaction.
}
\end{figure}

The EDM and the chromo-EDM of fermions seen above are P, CP-odd quantities generated in a straightforward way  by elementary processes of a single fermion. 
For the case of composite systems such as hadrons, nuclei and atoms on the other hand, their EDMs are produced through the P, CP-odd four-fermion interactions.
The trilinear RPV interactions (\ref{eq:rpvlagrangian}) contribute to the P, CP-odd four-fermion interaction at the tree level as shown in Fig. \ref{fig:4-fermion} \cite{herczege-n,faessler}.
The four-fermion interaction can be written as follows:
\begin{equation}
{\cal L}_{\rm 4f} 
=
\frac{{\rm Im} (\hat \lambda_{ijj} \tilde \lambda^*_{ikk})}{2 m_{\tilde \nu_i}^2} \left[
\bar f_j f_j \cdot \bar F_k i \gamma_5  F_k
-\bar f_j i \gamma_5 f_j \cdot \bar F_k F_k
\right] 
,
\label{eq:rpv4-fermioninteraction}
\end{equation}
where $\hat \lambda = \lambda$ or $ \lambda'$, depending on whether the fermion $f$ is a lepton or quark.
Also, $\tilde \lambda = \lambda$ or $ \lambda'$ when the fermion $F$ is a lepton or quark, respectively.

In this analysis, the sparticle mass (to be denoted generically by $m_{\rm SUSY}$) is taken to be $m_{\rm SUSY} > 1$ TeV, in accordance with the exclusion region of the LHC experiment \cite{susylhc}.
We also assume that the flavor off-diagonal terms of the soft SUSY breaking terms are small \cite{gabbiani}. 
We assume that the CP violation in the $R$-parity conserving sector is minimal. 
The $\theta$-term is removed with Peccei-Quinn symmetry \cite{peccei}. 
In addition, we assume that the sfermion masses are degenerate. 
Under this condition, the contribution from the 2-loop level rainbowlike diagram vanishes \cite{yamanakarainbow1,yamanakarainbow2}.

\subsection{\label{sec:classification}Classification of the RPV contribution}

As we see in Eqs.  (\ref{eq:bzfunction}) and 
(\ref{eq:rpv4-fermioninteraction}), fermion EDMs and CP-odd four-fermion 
interactions depend on imaginary parts of certain combinations of bilinear products of 
$\hat \lambda _{ijk}$ and/or $\tilde \lambda ^{\prime (*)}_{ink}$. The combinations depend on EDM-measured objects and it will be helpful to classify the bilinear products 
before we start detailed analyses. With these {\colourcyan{purposes}} in our mind, 
we have classified, in the previous work \cite{yamanakabook}, the RPV contributions to the EDMs into 6 types:
\begin{itemize}
\item{Type 1:}
Leptonic bilinears which contribute only to the electron EDM via the Barr-Zee diagram [Im($\lambda_{311}\lambda^*_{322}$) and Im($\lambda_{211}\lambda^*_{233}$)].
The EDM of paramagnetic atoms and molecules are very sensitive to them.

\item{Type 2:}
Semi-leptonic bilinears involving electron which contribute both to the electron EDM and P, CP-odd electron-nucleon (e-N) interactions [Im($\lambda_{i11}\lambda'^*_{i11}$), Im($\lambda_{i11}\lambda'^*_{i22}$) and Im($\lambda_{i11}\lambda'^*_{i33}$) ($i=2,3$)].
Atomic EDMs (paramagnetic and diamagnetic) are very sensitive to them.

\item{Type 3:}
Semi-leptonic bilinears involving d-quark and heavy leptons. These can be only constrained via nucleon EDM [Im($\lambda_{i22}\lambda'^*_{i11}$) ($i=1,3$) and Im($\lambda_{j33}\lambda'^*_{j11}$) ($j=1,2$)].

\item{Type 4:}
Hadronic bilinears. They contribute to the d-quark EDM, choromo-EDM and P, CP-odd 4-quark interactions [Im($\lambda'_{i11}\lambda'^*_{i22}$), Im($\lambda'_{i11}\lambda'^*_{i33}$) and Im($\lambda'_{i22}\lambda'^*_{i33}$) ($i=1,2,3$)].
Purely hadronic EDMs (nucleon EDMs, bare nuclear EDMs) are highly sensitive to them.

\item{Type 5:}
Bilinears which contribute only to muon EDM [Im($\lambda_{122}\lambda^*_{133}$), Im($\lambda_{i22}\lambda'^*_{i22}$) and Im($\lambda_{i22}\lambda'^*_{i33}$) ($i=1,3$)].

\item{Type 6:}
Remaining RPV bilinears which cannot be constrained in this analysis [Im($\lambda_{i33}\lambda'^*_{i22}$) and Im($\lambda_{i33}\lambda'^*_{i33}$) ($i=1,2$)].
They are expected to contribute to the EDMs of $\tau$ lepton, $s$ and $b$ quarks. 

\end{itemize}

We are well aware of experiments trying to 
measure the muon and $\tau$ lepton EDM \cite{pdg},
and they could provide us with stringent constraints in the near 
future \cite{raidal}. 
However for now,  we have not considered the Type 5 and Type 6 contributions
in our analysis, which can be studied independently from the atomic and 
nuclear EDM's.

\subsection{P, CP-odd electron-nucleon and pion-nucleon interactions}

In the present subsection, we would like to outline the derivation of the 
P- and CP-odd electron-nucleon and pion-nucleon interactions which are 
both indispensable for computation of nuclear and atomic EDMs 
in Section \ref{sec:many-body}.

Let us begin with the P, CP-odd electron-nucleon (e-N) interactions \cite{e-nint,khriplovichbook,yamanakabook} described by
\begin{eqnarray}
{\cal L}_{eN} 
&=&
 -\frac{G_F}{\sqrt{2}} \sum_{N=p,n} \Bigl[
C_N^{\rm SP} \bar NN \, \bar e i \gamma_5 e
+C_N^{\rm PS} \bar Ni\gamma_5 N \, \bar e e 
\nonumber\\
&& \hspace{8em}
+\frac{1}{2}C_N^{\rm T} \epsilon^{\mu \nu \rho \sigma} \bar N \sigma_{\mu \nu} N \, \bar e \sigma_{\rho \sigma} e 
\Bigr] .\ \ \ \ 
\label{eq:pcpve-nint}
\end{eqnarray}
These interactions originally come from 
the P, CP-odd electron-quark interactions with the same Lorentz structure, i.e., 
\begin{eqnarray}
{\cal L}_{eq} 
&=&
 -\frac{G_F}{\sqrt{2}} \sum_{q=d,s,b} \Bigl[ 
 C_q^{\rm SP} \bar qq \, \bar e i \gamma_5 e+C_q^{\rm PS} \bar q i\gamma_5 q \, \bar e e 
\nonumber\\
&& \hspace{8em}
+\frac{1}{2}C_q^{\rm T} \epsilon^{\mu \nu \rho \sigma} \bar q \sigma_{\mu \nu} q \, \bar e \sigma_{\rho \sigma} e 
\Bigr] .
\label{eq:electronquarkinteractions}
\end{eqnarray}
The coefficients $C_{q}^{\rm SP}$ and $C_{q}^{\rm PS}$ are 
obtained by looking at the r.h.s. of Eq. (\ref{eq:rpv4-fermioninteraction}).
The coefficients $C_{N}^{\rm SP}$, $C_{N}^{\rm PS}$, and $C_{N}^{\rm T}$ in (\ref{eq:pcpve-nint}) are extracted simply by 
attaching  a factor of the quark content of the nucleon to 
the corresponding ones, i.e., $C_{q}^{\rm SP}$, $C_{q}^{\rm PS}$, and $C_{q}^{\rm T}$ in Eq. (\ref{eq:electronquarkinteractions}).
We are thus led to the formulae
\begin{eqnarray}
C_N^{\rm SP} \bar N N 
&=&
C_q^{\rm SP} \langle N| \bar q q | N \rangle
,
\label{eq:cnsp} 
\\
C_N^{\rm PS} \bar N i\gamma_5 N 
&=&
C_q^{\rm PS} \langle N| \bar qi\gamma_5 q | N \rangle
,
\label{eq:cnps} 
\\
C_N^{\rm T} \bar N \sigma_{\mu \nu} N
&=&
C_q^{\rm T} \langle N| \bar q \sigma_{\mu \nu} q | N \rangle
.
\label{eq:cnt}
\end{eqnarray}
The detail of the nucleon matrix elements will be given below.
The tensor-type P, CP-odd e-N interaction of Eq. (\ref{eq:pcpve-nint}) does not receive the RPV contribution at the level of our discussion.
It is however useful to mention it since some P, CP-odd atomic level effects can be calculated using the tensor-type P, CP-odd e-N interaction (see Section \ref{sec:many-body}).

Next let us turn to the P, CP-odd pion-nucleon interactions described effectively by 
\begin{eqnarray}
{\cal L}_{\pi NN} 
&=& 
\sum _{N=p,n} \sum_{a=1}^3 \Big[ \bar g_{\pi NN}^{(0)} \bar N \tau^a N \pi^a 
+ \bar g_{\pi NN}^{(1)} \bar N N \pi^0
\nonumber  \\
& & 
+ {\bar g}^{(2)}_{\pi NN}\left (
\Bar N\tau ^{a} N \pi ^{a}-3\bar N \tau ^{3} N \pi ^{0}
\right )
\Big] \, ,
\label{eq:pcpvpinnint}
\end{eqnarray}
where $a$ denotes the isospin index.
The third term in Eq. (\ref{eq:pcpvpinnint})
is of tensor-type and will not be discussed because its effect is negligibly small. 
The first two terms with coefficients $\bar g_{\pi NN}^{(0)} $ and 
$\bar g_{\pi NN}^{(1)} $, respectively receive two types of contributions. One is 
due to the quark chromo-EDM $d_{q}^{c}$ 
and the other is due to P, CP-odd four-quark 
interactions. As to the former we do not know available lattice QCD data.
In this work, we use therefore the result of the QCD sum rules \cite{qcdsumrules,pospelovreview}. 
The contributions of the quark chromo-EDM to $\bar g_{\pi NN}^{(0)} $ and $\bar g_{\pi NN}^{(1)} $ have been shown \cite{pospelovreview,pospelov} as 
\begin{eqnarray}
\bar g_{\pi NN}^{(0)} (d^c_q)
&=&
\tilde \omega_{(0)} \frac{d^c_u + d_d^c}{10^{-26}{\rm cm}}
,
\\
\bar g_{\pi NN}^{(1)} (d^c_q)
&=&
\tilde \omega_{(1)} \frac{d^c_u - d_d^c}{10^{-26}{\rm cm}}
,
\end{eqnarray}
where
\begin{eqnarray}
\tilde \omega_{(0)} 
&=&
0.95\times 10^{-12} \times \frac{|\langle 0 |\bar qq|0\rangle | }{(262\, {\rm MeV})^3}
\frac{|m_0^2|}{0.8 \, {\rm GeV}^2}
,
\\
\tilde \omega_{(1)} 
&=&
4.7\times 10^{-12} \times \frac{|\langle 0 |\bar qq|0\rangle | }{(262\, {\rm MeV})^3}
\frac{|m_0^2|}{0.8 \, {\rm GeV}^2}
.
\end{eqnarray}
Here and hereafter we set \cite{belayev}
\begin{eqnarray}
\langle 0 |\bar qq|0\rangle 
&=&
 -(262\, {\rm MeV})^3, 
\nonumber \\
m_0^2 
&\equiv &
\langle 0 |\bar q g_s \sigma_{\mu \nu} G^{\mu \nu}_a t_a q|0\rangle / \langle 0 |\bar qq|0\rangle
\nonumber \\
&=&
 -(0.8\pm 0.1) \, {\rm GeV}^2\: .
\end{eqnarray}

The effect of the P, CP-odd four-quark interaction to $\bar g_{\pi NN}^{(1)} $ on the other hand is given by the factorization approximation \cite{pospelovreview,yamanakabook,4-quark}:
\begin{eqnarray}
\bar g_{\pi NN}^{(1)} (C_q) 
&=& 
\sum_{q = s,b} \langle \pi^0 N | C_q \,  \bar q q  \, \bar d i\gamma_5 d | N \rangle \nonumber\\
&\approx&
-\frac{F_\pi m_\pi^2 }{2m_d} \sum_{q = s,b} C_q \langle N | \, \bar q q | N \rangle 
,
\label{eq:barg1pinncq}
\end{eqnarray}
where the P, CP-odd four-quark coupling $C_q$ has to be matched with the coefficient of Eq. (\ref{eq:rpv4-fermioninteraction}).
Here we also need the data of the scalar content of nucleon.
We should note that in this paper, only the isovector P, CP-odd pion-nucleon interaction receives contribution from the P, CP-odd four-quark interactions.

We are now in a position to present the detailed quark contents in nucleons which are necessary to evaluate (\ref{eq:cnsp}), (\ref{eq:cnps}) and (\ref{eq:barg1pinncq}) and the EDM of the composite system.
The hadron level calculation, which is the integral part in the EDM predictions, has some subtleties, and we have to explain it in detail.
It is most favorable that the hadronic matrix elements are given by the lattice QCD calculation \cite{yamanakabook,bhattacharya}.
In our calculation, we have used the lattice QCD result for the quark scalar contents of nucleon \cite{lattice,lattice_charm_content,bhattacharya2}.
We use the following nucleon matrix elements renormalized at $\mu =2$ GeV:
\begin{eqnarray}
\langle p| \bar uu |p\rangle &=& 6.9 
,
\label{eq:uu}
\\
\langle p| \bar dd |p\rangle &=& 5.9 
,
\label{eq:dd}
\\
\langle p| \bar ss |p\rangle &=& 0.1 .
\label{eq:ss}
\end{eqnarray}
For the up and down contents, we have set the value of the nucleon sigma term 
\begin{equation}
\frac{m_u +m_d}{2} \langle p| \bar uu + \bar dd |p\rangle \approx 45\:\:{\rm  MeV} 
,
\end{equation}
favored by lattice QCD studies \cite{lattice} and the isovector content $\langle p| \bar uu - \bar dd |p\rangle \approx 1.02$, also derived from the analysis of several lattice QCD results \cite{alonso,rpvbetadecay}.
The strange quark content was given by the lattice QCD studies \cite{lattice,lattice_charm_content}.
The quark masses that we use are \cite{pdg}
 \begin{eqnarray}
 m_u &=& 2.2 \:\:{\rm MeV},
 \\ 
 m_d &=& 4.8 \:\:{\rm MeV}, 
\\
 m_s &=& 100  \:\:{\rm  MeV}\:\:.
 \end{eqnarray}
 
The scalar density of the neutron $\langle n| \bar qq |n \rangle$ is also necessary when we compute (\ref{eq:cnsp}), (\ref{eq:cnps}) and (\ref{eq:barg1pinncq}).
To obtain the neutron matrix elements, we simply use the isospin symmetry, i.e. $\langle n| \bar dd |n\rangle = \langle p| \bar uu |p\rangle$ ($\bar uu \leftrightarrow \bar dd$), and $\langle n| \bar ss |n\rangle = \langle p| \bar ss |p\rangle$.

Another lattice QCD result we quote is the tensor content of the nucleon \cite{latticetensorcharge,bhattacharya2}.
The proton tensor charge is expressed in terms of the momentum $p^\mu$ and the proton spin $s^\mu$ by
\begin{equation}
\langle p (p,s) | \bar q i\sigma^{\mu \nu} \gamma_5 q | p (p,s) \rangle
=
2 (s^\mu p^\nu -s^\nu p^\mu ) 
\delta q
,
\end{equation}
where $\delta q$ ($q=u,d,s$) is the tensor content (charge) of the proton.
The tensor charge of the nucleon gives the linear coefficients of the contribution of the quark EDM to the nucleon EDM \cite{bhattacharya,yamanakabook}
\begin{equation}
d_p (d_q) \bar N \sigma^{\mu \nu} N
=
d_q \langle p | \bar q \sigma^{\mu \nu} q | p \rangle
=
\delta q \,
d_q \bar N \sigma^{\mu \nu} N
,
\end{equation}
and also the linear coefficient of the contribution of the tensor-type P, CP-odd electron-quark interaction to the P, CP-odd e-N interaction \cite{khriplovichbook} [see Eq. (\ref{eq:cnt})].
The lattice QCD result of the proton tensor charge is \cite{latticetensorcharge}
\begin{eqnarray}
\delta u &=& 0.8 ,\\
\delta d &=& -0.2 ,\\
\delta s &=& -0.05 .
\end{eqnarray}
By using the isospin symmetry, we have 
\begin{eqnarray}
\langle p | \bar u \sigma^{\mu \nu} u | p\rangle 
= \langle n | \bar d \sigma^{\mu \nu} d | n\rangle,
\\ 
\langle n | \bar u \sigma^{\mu \nu} u | n\rangle 
= \langle p | \bar d \sigma^{\mu \nu} d | p\rangle,
\\ 
\langle n | \bar s \sigma^{\mu \nu} s | n\rangle 
= \langle p | \bar s \sigma^{\mu \nu} s | p\rangle \:\:.
\end{eqnarray}
It should be noted that the tensor charge obtained from lattice QCD is smaller than the nonrelativistic quark model predictions 
$\delta u 
= \frac{4}{3}$ and $\delta d
= -\frac{1}{3}$, often used in the literature \cite{adler}.
The small values of the tensor charge is explained by the superposition of the processes in which the gluon emission and absorption of the quark flip the quark tensor charge \cite{tensorsde}.

The pseudoscalar contents of nucleon have been calculated phenomenologically, as \cite{cheng,herczege-n,yamanakabook,alonso}
\begin{eqnarray}
\langle p| \bar ui\gamma_5 u |p\rangle &=& 173  
,
\label{eq:u5u}
\\
\langle p| \bar di\gamma_5 d |p\rangle &=& -134 
,
\label{eq:d5d}
\\
\langle p| \bar si\gamma_5 s |p\rangle &=& -3.0  
, 
\label{eq:s5s}
\end{eqnarray}
where the recent experimental data of the nucleon axial charge were used as input \cite{compass,ucna}.
The large value of the pseudoscalar condensates for the light quarks is due to the pion pole contribution \cite{axialsde}.
For the detailed derivation, see Appendix \ref{sec:pseudoscalar}.

The scalar and pseudoscalar bottom contents of the nucleon can be obtained via the heavy quark expansion \cite{zhitnitsky}.
In this work, we use
\begin{eqnarray}
\langle p| \bar bb |p\rangle
&\approx & 
1.2\times 10^{-2}
,
\\
\langle p| \bar bi\gamma_5 b |p\rangle &=& -6.8 \times 10^{-2}
.
\label{eq:b5b}
\end{eqnarray}
Here we also assume the isospin symmetry, i.e. $\langle n| \bar bb |n\rangle = \langle p| \bar bb |p\rangle$ and $\langle n| \bar bi\gamma_5b |n\rangle = \langle p| \bar bi\gamma_5b |p\rangle$.
See Appendix \ref{sec:heavy_quark} for the derivation.

\subsection{The chromo-EDM and the nucleon EDM}

We now evaluate the chromo-EDM contribution [see Eq. (\ref{eq:cedm})] to the nucleon EDM \cite{gunion}.
In this case again we do not have any lattice QCD results at our disposal, but there are many model calculations \cite{gunion,chemtob,falk,pospelovchromoedm,hisano,faessler,hisanochromoedm,fuyuto,yamanakabook,chiraledm}, despite the large theoretical uncertainty.
As the two main ways often used to discuss the chromo-EDM contribution to the nucleon EDM, we know the chiral and QCD sum rules approaches \cite{pospelovreview}.
The first approach is the traditional one, which has its origin in the investigation of the observability of the $\theta$-term effect \cite{peccei,crewther,pich,chiraledm}.
There it was argued that the leading contribution of the $\theta$-term to the nucleon EDM is given by the pion cloud effect.
The chromo-EDM is known to contribute to the nucleon EDM in a similar way.
The second approach is a more recent one, focusing on the elementary level quark-gluon processes \cite{pospelovchromoedm,hisanochromoedm}.

\begin{figure}[htb]
\includegraphics[width=9cm]{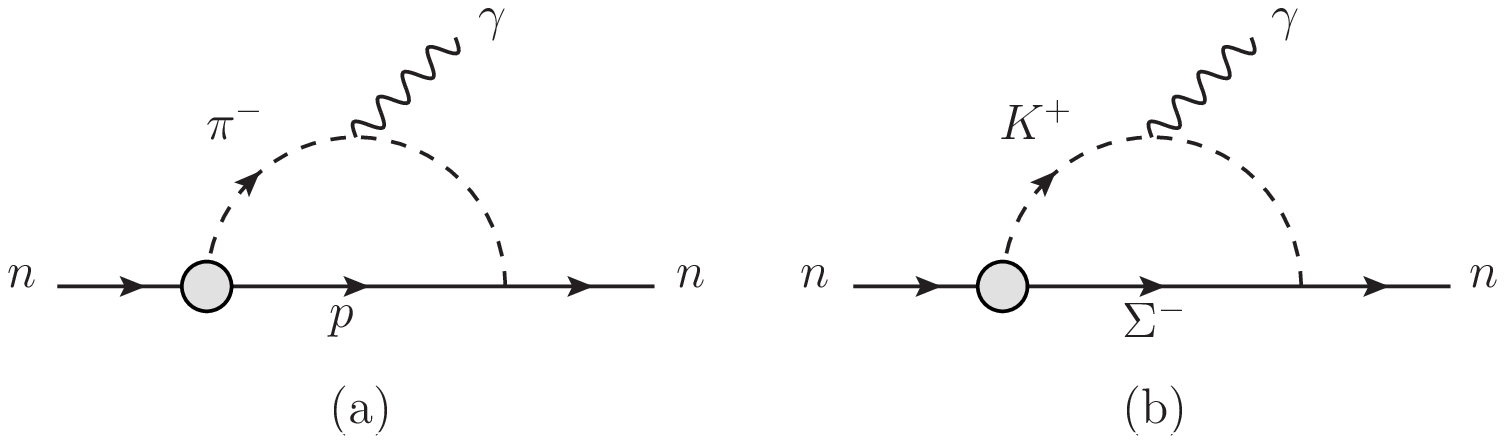}
\caption{\label{fig:neutronmesonloop} 
Nambu-Goldstone boson loop contribution to the neutron EDM.
The grey blob represents the P, CP-odd interaction.}
\end{figure}

In this work, we consider the chiral approach to highlight physical contents
of the nucleon EDM.
The reason is as follows.
In the phenomenological determination of the pseudoscalar contents of nucleon (\ref{eq:u5u}), (\ref{eq:d5d}), and (\ref{eq:s5s}), we have pointed out the importance of the pion pole contribution which gives very large matrix elements \cite{axialsde}.
In a similar way, it is natural to consider the Nambu-Goldstone boson contribution in this case, where the nucleon EDM is enhanced by a chiral logarithm.
The result of the calculation actually yields larger coefficients than those obtained from the QCD sum rules analysis \cite{fuyuto,yamanakabook}.
The recent investigations in the chiral approach are pursued by using the chiral effective theory \cite{chiraledm}.
In this work, we have quoted the result {\colourcyan{of}} Ref. \cite{fuyuto}.
The contribution of the quark chromo-EDM to the neutron EDM is given by
\begin{equation}
d_n (d_q^c) =
e\tilde \rho_n^u d_u^c
+e\tilde \rho_n^d d_d^c
+e\tilde \rho_n^s d_s^c
,
\end{equation}
where $\tilde \rho_n^u \approx 3.0$, $\tilde \rho_n^d \approx 2.5$, and $\tilde \rho_n^s \approx 0.5$.
This result was obtained by taking the leading chiral logarithm of the Nambu-Goldstone boson loop diagram as shown in Fig. \ref{fig:neutronmesonloop}.
The Peccei-Quinn symmetry \cite{peccei} was assumed, so that the quark chromo-EDM contributes also through the $\theta$-term.
In this derivation, the P, CP-odd pion-nucleon coupling was evaluated in the QCD sum rules.

\begin{figure}[htb]
\includegraphics[width=9cm]{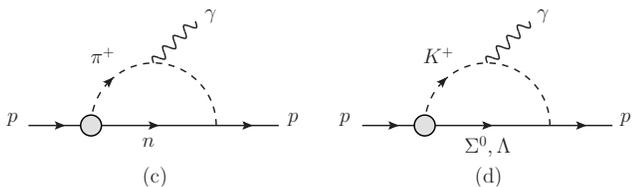}
\caption{\label{fig:protonmesonloop} 
Nambu-Goldstone boson loop contribution to the proton EDM.
The grey blob represents the P, CP-odd interaction.}
\end{figure}

In our analysis, the expression of the proton EDM is also needed.
The calculation of the proton EDM can be done in a similar way, by evaluating Fig. \ref{fig:protonmesonloop} \cite{yamanakabook}.
At the leading order of chiral logarithm, the convenient relations $\tilde \rho_p^d=-\tilde \rho_n^d$ and $\tilde \rho_p^u = \tilde \rho_p^d + \tilde \rho_p^s$ hold.
Moreover, at this order the coefficient $\tilde \rho_p^s$ is related to the $\tilde \rho_n^s$ by 
\begin{equation}
\frac{\tilde \rho_p^s}{\tilde \rho_n^s}
=
\frac{\frac{D-F}{2}(S_d -S_s ) +\frac{D+3F}{6}(S_d +S_s -2S_u)}{(D-F) (S_d -S_s)}
,
\label{eq:rhosp_rhosn}
\end{equation}
where $S_q \equiv \langle p| \bar qq |p\rangle \ (q=u,d,s)$, $D=0.80$ and $F=0.47$ (see Appendix \ref{sec:appendixnucleonedm} for details).
If we use the input parameters of Eqs. (\ref{eq:uu}), (\ref{eq:dd}), and (\ref{eq:ss}), we obtain $\tilde \rho_p^s \approx -1.0 \times \tilde \rho_n^s$, so that $\tilde \rho_p^u \approx -\tilde \rho_n^u$.
The coincidence of these two relations should be considered as accidental.
We should note that the renormalization scale of Ref. \cite{fuyuto} ($\mu =1$ GeV) is different from that of this paper ($\mu =2$ GeV).
This difference should not be considered seriously as the theoretical uncertainty is large. 
If we express the proton EDM explicitly, we have
\begin{equation}
d_p (d_q^c) =
e\tilde \rho_p^u d_u^c
+e\tilde \rho_p^d d_d^c
+e\tilde \rho_p^s d_s^c
,
\end{equation}
where $\tilde \rho_p^u \approx -3.0$, $\tilde \rho_p^d \approx -2.5$, and $\tilde \rho_p^s \approx -0.5$.
The important feature of the results of the chiral approach is that the strange quark chromo-EDM contribution is not small.


In this work, we have neglected the effect of the QCD renormalization, which would give a mixing between P, CP-odd quark level operators under the change of the renormalization scale \cite{degrassi,renormalizationedm}.
It should be noted in particular that the effect of the Weinberg operator ${\cal L}_w = \frac{1}{6} w \frac{G_F}{\sqrt{2}} f^{abc} \epsilon^{\alpha \beta \gamma \delta} G^a_{\mu \alpha } G_{\beta \gamma}^b G_{\delta}^{\mu,c}$ \cite{weinbergoperator,chemtob} appears from the renormalization, although the RPV interactions do not generate it at the leading order.
The hadron matrix elements calculated in this work however involve a large theoretical uncertainty, and the mixing effect should be well within it.
We therefore neglect the renormalization of the operators together with the Weinberg operator.

\subsection{\label{sec:many-body}Many-body physics}

To determine the EDM of the nuclear and atomic level many-body systems, we need to calculate their wave functions.
The leading P, CP-odd contribution used as input is the P, CP-odd pion-nucleon interactions [see Eq. (\ref{eq:pcpvpinnint})] and the nucleon EDM.
The linear coefficients in the relation between these P, CP-odd nucleon level interactions to the nuclear EDM has to be worked out.
For the calculation of the few-body systems such as the deuteron or the $^3$He nucleus, the ab initio approach is possible and in fact turns out to be effective \cite{liu,gudkov,stetcu}.
On the other hand, the wave functions of the systems with more than $O(100)$-body cannot be calculated without any approximations.

For heavy nuclei in atoms, the P, CP-odd effect is suppressed by the screening phenomenon of Schiff \cite{schiff}, but the finite volume effect can generate the atomic EDM through the nuclear Schiff moment \cite{ginges}.
In the evaluation of the nuclear Schiff moment, we also calculate the linear coefficients between the nuclear Schiff moment and the P, CP-odd pion-nucleon interactions [see Eq. (\ref{eq:pcpvpinnint})] or the nucleon EDM.
The general expression of the nuclear Schiff moment of the nucleus $a$ is given as
\begin{eqnarray}
S_a 
&=&
s_p^a d_p +s_n^a d_n
\nonumber\\
&&
+ g_{\pi NN } \Bigl[ a^{(0)}_a g^{(0)}_{\pi NN} +a^{(1)}_a g^{(1)}_{\pi NN} +a^{(2)}_a g^{(2)}_{\pi NN} ) \Bigr]
,
\nonumber\\
\label{eq:nuclearschiffmoment}
\end{eqnarray}
where $g^{(0)}_{\pi NN} $, $g^{(1)}_{\pi NN} $, and $g^{(2)}_{\pi NN} $ are the isoscalar, isovector, and isotensor P, CP-odd pion-nucleon couplings, respectively.
The CP-even pion-nucleon coupling is given by $g_{\pi NN} =14.11 \pm 0.20$ \cite{ericson}.

For the calculations of the coefficients of $a_{a}^{(0)}$, $a_{a}^{(1)}$ and $a_{a}^{(2)}$ in Eq. (\ref{eq:nuclearschiffmoment}) for $^{129}$Xe nucleus, we use the result of the shell model \cite{yoshinaga1,yoshinaga2}.
For the calculations of the nuclear Schiff moments of 
the $^{199}$Hg, $^{211}$Rn, and $^{225}$Ra, we use the results of the many-body calculation with mean-field approximation taking into account the deformation \cite{ban,dobaczewski}.
There are currently variety of phenomenological interactions available.
Calculations were made using the computer code HFODD \cite{hfodd} within several models of phenomenological Skyrme interactions: SkO' \cite{sko}, SkM$^*$ \cite{skm}, SLy4 \cite{sly4}, SV \cite{sv} and SIII \cite{sv}. 
The results for $^{199}$Hg, $^{211}$Rn, and $^{225}$Ra nuclei are shown in Table \ref{table:schiffmoment}.

In Ref. \cite{ban}, the dependence of the $^{199}$Hg and $^{211}$Rn nuclear Schiff moments on the valence neutron EDM, namely the coefficients $s_{n}^{a}$ in (\ref{eq:nuclearschiffmoment}), is  given by the parameter $b_a$ which can be converted to $s_n^a$ owing to the formula \cite{ban}
\begin{equation}
s_n^a
=
-b_a 
\frac{4\pi^2 m_N}{e \ln (m_N / m_\pi ) } \cdot \frac{ \langle \sigma^a_n \rangle}{ \langle \sigma^a_N \rangle}
.
\label{eq:s_n}
\end{equation}
To obtain the coefficient  $s_p^a$ of the proton EDM dependence in Eq. (\ref{eq:nuclearschiffmoment}), we have to replace the nucleon spin matrix element as $\langle \sigma^a_n \rangle \rightarrow \langle \sigma^a_p \rangle$ in Eq. (\ref{eq:s_n}).
The nuclear spin matrix elements will be calculated below.
For the $^{199}$Hg and $^{211}$Rn nuclei, we take the average of the coefficients calculated with different phenomenological interactions \cite{ban}.
For the $^{225}$Ra nucleus, we take the result obtained with the SkO' interaction, as recommended in Ref. \cite{dobaczewski}.
We see that the results have a large theoretical uncertainty.

\begin{table}
\caption{Coefficients $a^{(i)}$ of the dependence of the Schiff moment on P, CP-odd pion-nucleon couplings ($S_a = g_{\pi NN} (a_a^{(0)} \bar g^{(0)}_{\pi NN} +a_a^{(1)} \bar g^{(1)}_{\pi NN} + a_a^{(2)} \bar g^{(2)}_{\pi NN})$) in unit of $e$ fm$^3$ for the $^{199}$Hg, $^{211}$Rn, and $^{225}$Ra atoms, calculated with different phenomenological interactions. 
The labels HB and HFB stand for calculations in the Hartree-Fock and Hartree-Fock-Bogoliubov approximations, respectively.}
\begin{ruledtabular}
\begin{tabular}{lcccc}
$^{199}$Hg& $-a^{(0)}_{\rm Hg}$\ \  &$-a^{(1)}_{\rm Hg}$ \ \ &$a^{(2)}_{\rm Hg}$\ \ &$-b_{\rm Hg}$ \ \  \\ 
\hline
SkM$^*$ (HFB) & 0.041 & $-$0.027 & 0.069 & 0.013 \\
SLy4 (HFB)& 0.013 & $-$0.006 & 0.024 & 0.007 \\
SLy4 (HF)& 0.013 & $-$0.006 & 0.022 & 0.003 \\
SV (HF)& 0.009 & $-$0.0001 & 0.016 & 0.002 \\
SIII (HF)& 0.012 & 0.005 & 0.016 & 0.004 \\
\hline
Average&0.018&$-$0.007 &0.029&0.0058\\
\end{tabular}
\end{ruledtabular}

\begin{ruledtabular}
\begin{tabular}{lcccc}
$^{211}$Rn& $-a^{(0)}_{\rm Rn}$\ \  &$-a^{(1)}_{\rm Rn}$ \ \ &$a^{(2)}_{\rm Rn}$\ \ &$-b_{\rm Rn}$ \ \  \\ 
\hline
SkM$^*$ & 0.042 & $-$0.028 & 0.078 & 0.015 \\
SLy4 & 0.042 & $-$0.018 & 0.071 & 0.016 \\
SIII & 0.034 & $-$0.0004 & 0.064 & 0.015 \\
\hline
Average&0.039&$-$0.0015 &0.071 &0.0015\\
\end{tabular}
\end{ruledtabular}

\begin{ruledtabular}
\begin{tabular}{lcccc}
$^{225}$Ra& $-a^{(0)}_{\rm Ra}$\ \  &$-a^{(1)}_{\rm Ra}$ \ \ &$a^{(2)}_{\rm Ra}$\ \ &$-b_{\rm Ra}$ \ \  \\ 
\hline
SkM$^*$ & 4.7 & $-$21.5 & $-$11.0 & $-$ \\
SLy4 & 3.0 & $-$16.9 & $-$8.8 & $-$ \\
SIII & 1.0 & $-$7.0 & $-$3.9 & $-$ \\
\hline
SkO' &1.0&$-$6.0 &$-$4.0 &$-$\\
\end{tabular}
\end{ruledtabular}

\label{table:schiffmoment}
\end{table}

We also introduce the way to calculate the spin matrix elements of the valence proton $\langle \sigma_p \rangle$ and neutron and $\langle \sigma_n \rangle$ using the nuclear magnetic moment.
These elements are needed to separate the CP-odd effect for the valence nucleon EDM contribution to the nuclear Schiff moment and for the pseudoscalar-type P, CP-odd e-N interaction $C^{\rm PS}_N$ [see Eq. (\ref{eq:pcpve-nint})].
Phenomenologically, the valence nucleon is a superposition of the proton and neutron due to the configuration mixing.
The mixing coefficients $\langle \sigma_p \rangle$ and $\langle \sigma_n \rangle$ can be obtained using the magnetic moment of the nucleus $a$ as follows:
\begin{equation}
\left\{
\begin{array}{ccl}
\mu_a
&= & 
\mu_p \langle \sigma_p^a \rangle + \mu_n \langle \sigma_n^a \rangle  
, \\
\langle  \sigma^a_N \rangle
&=& 
\langle \sigma^a_n \rangle + \langle \sigma^a_p \rangle 
\,
\end{array}
\right.
\label{eq:pnmixingcoefficient}
\end{equation}
where $\mu_a$ is the nuclear magnetic moment (in unit of the nuclear magneton).
Here the matrix element 
$\langle \sigma^a_N \rangle$ is 1 for $j=l+\frac{1}{2}$ nuclei, and $-\frac{j}{j+1}$ for $j=l-\frac{1}{2}$ nuclei.
The magnetic moment of the proton is $\mu_p = +2.7928$ and that of the neutron is $\mu_n = -1.9130$.

The final matter is the EDMs of atoms and molecules.
In the atomic level evaluation of the EDM, we calculate the linear coefficients between the atomic (or molecular) EDMs and the nuclear Schiff moment, the P, CP-odd e-N interactions [see Eq. (\ref{eq:pcpve-nint})] or the electron EDM.

For the paramagnetic systems such as the $^{205}$Tl, $^{210}$Fr atoms, YbF and ThO molecules, the enhancement factor of the electron EDM $K$ and the scalar-type P, CP-odd electron-nucleon interaction $C^{\rm SP}_N$ [see Eq. (\ref{eq:pcpve-nint})] are calculated directly using the Relativistic Hartree-Fock approach \cite{flambaumtl,flambaumfr,flambaumybftho}.
The EDM of the paramagnetic atom or molecule is expressed as
\begin{equation}
d_a = K^a \left[ d_e + R^a_{\rm SP} \left( \frac{Z}{A} C^{\rm SP}_p + \frac{N}{A} C^{\rm SP}_n \right)
\right]
,
\end{equation}
where $Z$, $N$, and $A$ are the proton, neutron and the total nucleon numbers of the nucleus $a$, respectively.
For the case of the paramagnetic molecule, $Z$, $N$ and $A$ are those of the biggest nucleus.
The first term is the contribution of the electron EDM, with the enhancement factor $K^a$, and the second term that from the scalar-type P, CP-odd e-N interactions.
The effect of the P, CP-odd e-N interactions was expressed relative to the electron EDM contribution, because the experimental data of the paramagnetic systems are often written in terms of the electron EDM.
In this work, we neglect the contribution of the pseudoscalar-type P, CP-odd e-N interaction and the nuclear Schiff moment for the $^{205}$Tl atoms and the YbF molecule.
The theoretical uncertainty of the input parameters for the paramagnetic systems is known to be small, within few percents, and the results given by other approach \cite{kelly,porsev,mukherjee,ybfkozlov,nayak} are more or less consistent.

The EDM of diamagnetic atoms has a moderate sensitivity on the electron EDM or the scalar-type P, CP-odd e-N interaction, due to the closed electron shell.
It is therefore important to also consider the effect of other P, CP-odd effects such as the nuclear Schiff moment or the pseudoscalar-type P, CP-odd e-N interaction.
In this work, the EDM of the diamagnetic atoms is expressed as
\begin{eqnarray}
d_a 
&=&
K^a_{d_e} d_e 
\nonumber\\
&&
+ K^a_{\rm SP} \left( \frac{Z}{A} C^{\rm SP}_p + \frac{N}{A} C^{\rm SP}_n \right)
\nonumber\\
&&
+ K^a_{\rm PS} \left< C^{\rm PS}_p \sigma_p^a + C^{\rm PS}_n \sigma_n^a \right>
\nonumber\\
&&
+ K^a_{\rm T} \left< C^{\rm T}_p \sigma_p^a + C^{\rm T}_n \sigma_n^a \right>
\nonumber\\
&&
+ K^S_a S_a 
,
\label{eq:atomiccoef}
\end{eqnarray}
where $\langle \sigma_N^a \rangle$ is the nuclear spin matrix element of the nucleus $a$, which can be calculated using Eq. (\ref{eq:pnmixingcoefficient}).
For the diamagnetic atoms relevant in this work ($^{129}$Xe, $^{199}$Hg, $^{211}$Rn, and $^{225}$Ra), the contributions from the nuclear Schiff moment, the pseudoscalar-type and the tensor-type P, CP-odd electron-nucleon interactions are calculated directly.
We have quoted the results of the Relativistic Hartree-Fock approach improved with random phase approximation or configuration interaction plus many-body perturbation theory \cite{flambaumdiamagnetic}.
The theoretical uncertainty of the input parameters for the diamagnetic atoms is also small, and the results given by other approach \cite{lathaxerayb} are almost consistent.
The detail of the parameters used in this paper is presented in the formula of the Section \ref{sec:setup}.

The other contributions such as the effects of the electron EDM or the scalar-type P, CP-odd e-N interactions are derived from the approximate analytic formulae in terms of the tensor-type P, CP-odd e-N interactions \cite{flambaumformula}.
The contribution of the electron EDM and the scalar-type P, CP-odd e-N interaction to the EDM of diamagnetic atoms can be analytically related to the effect of the tensor-type P, CP-odd e-N interaction.
The electron EDM contribution in diamagnetic atoms can be written as \cite{ginges,flambaumformula}
\begin{equation}
d_e  \frac{\vec I}{I} 
\leftrightarrow 
\frac{3}{7} \frac{G_F m_p e}{\sqrt{2} \pi \alpha_{\rm em} \mu_a } \frac{R}{R-1}
\left< C_p^{\rm T} \vec \sigma_p^a +C_n^{\rm T} \vec \sigma_n^a \right> \ ,
\label{eq:dect}
\end{equation}
where $R=\left( \frac{2}{\Gamma (2\gamma + 1)} \left( \frac{a_B}{2Zr} \right)^{1-\gamma} \right)^2 $ ($a_B$ is the Bohr radius), $\vec I$ the spin of the nucleus $a$, and $\mu_a$ the nuclear magnetic moment in unit of the nuclear magneton.
The scalar-type P, CP-odd e-N interaction $C_N^{\rm SP}$ can be written as \cite{ginges,flambaumformula}
\begin{eqnarray}
&&
\Biggl[ \frac{Z}{A} C_p^{\rm SP} + \frac{N}{A} C_n^{\rm SP} \Biggr] \frac{\vec I}{I}
\nonumber\\
&\leftrightarrow &
 \frac{1.9 \times 10^3 }{ (1+0.3Z^2\alpha_{\rm em}^2)A^{2/3} \mu_a } \left< C_p^{\rm T} \vec \sigma_p^a +C_n^{\rm T} \vec \sigma_n^a \right> 
 .
\label{eq:cspct}
\end{eqnarray}
The above two relations are accurate to $O( Z^2 \alpha_{\rm em}^2)$. 
For the diamagnetic atom EDM, the coefficient of the tensor-type P, CP-odd e-N interaction $ K^a_{\rm T}$ [see Table \ref{table:diamagnetic}] is often calculated.
The atomic EDM generated from the electron EDM and the scalar-type P, CP-odd e-N interaction are, respectively,
\begin{eqnarray}
d_a (d_e) 
&=&
d_e
K^a_{\rm T}
\frac{7\sqrt{2} \pi \alpha_{\rm em} \mu_a }{3 G_F m_p e} \frac{R-1}{R}
,
\\
d_a (C_N^{\rm SP} ) 
&=&
\Biggl[ \frac{Z}{A} C_p^{\rm SP} + \frac{N}{A} C_n^{\rm SP} \Biggr]
K^a_{\rm T}
\nonumber\\
&& \hspace{4em} \times
\frac{ (1+0.3Z^2\alpha_{\rm em}^2)A^{2/3} \mu_a }{1.9 \times 10^3 } 
.
\end{eqnarray}

\section{\label{sec:linearprogramming}The linear programming method}

In phenomenological perturbative analysis in general, we often encounter with constraint relations linear in physical observables.
To derive the maximum of some value (relation) constrained by these relations, we have to solve the set of inequalities, and find the maximum in the allowed region.
To do this, we can do a naive scan of the full parameter space by either dotting and checking or by either yes or no, the discretized space which is included in the
regions constrained by input inequalities.
However, this naive method looses efficiency when the parameter space becomes large, in particular when the dimension increases in number.

\begin{figure}[h]
\begin{center}
\includegraphics[height=56mm]{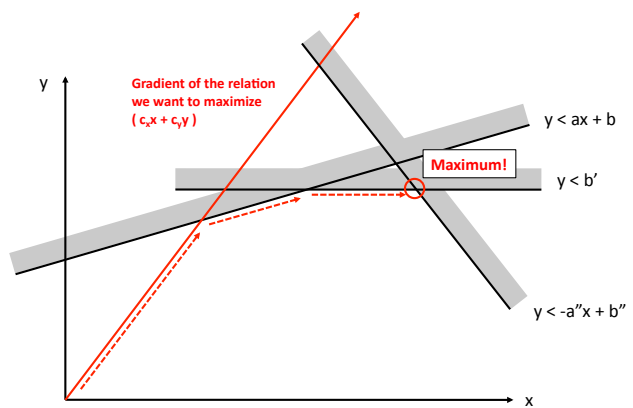}
\caption{\label{fig:linearprogramming} 
Schematic illustration of linear programming problem in two-dimension \cite{yamanakabook}.
}
\end{center}
\end{figure}

An efficient way to derive the maximum is the {\it linear programming method}.
This method is based on the observation that the maximum is located at one of the corners of the multidimensional polygon made out of inequality constraints (if the solution exists).
This can be understood as follows.
The linear relation we want to maximize constitutes a constant gradient.
Starting from somewhere in the allowed region, we follow the direction of the gradient to increase the linear relation.
When we reach one of the ``wall" (hyperplane) of constraint inequality, we follow then the direction of the projection of the gradient onto the wall.
The dimension of the hyperplane we hit in going along the projected gradient diminishes in turn, and we arrive finally at some of the corners of the multidimensional polygonal allowed region.
This is the point where the linear relation is maximized (of course if the gradient is found to be orthogonal with the final ``hyperwall" we strike in going along the projected gradient, the whole hyperwall will be a degenerate solution of the problem).
This is exactly the algorithm of the linear programming.
The schematic picture of the 2-dimensional example is shown in Fig. \ref{fig:linearprogramming}.
In our case, we want to predict the possible maximal value of the prepared experimental observables within the EDM-constraints.
The linear inequalities and observables must be expressed by variables, which are the combination of the RPV bilinears.

\section{\label{sec:setup}Calculational Procedures}

Let us now define all the input inequalities.
We first define the relevant degrees of freedom which are the imaginary parts of the bilinear of RPV couplings.
We then give the phenomenological upper bounds to the RPV couplings due to the known experimental data.
We finally present the linear relations between the RPV bilinears and the CP-odd observables in questions, which are needed to yield constraints.

\subsection{RPV degrees of freedom}

Let us define the set of RPV degrees of freedom in the linear programming.
It is adequate to choose linearly independent combinations of RPV bilinears that appear in the relation of EDMs as our variables.
The type 5 and 6 can be omitted from our analysis, since they are not related at all with the other observables.
We then obtain ten variables which span the RPV parameter space relevant in our discussion.
We define them as
\begin{eqnarray}
x_1&=& {\rm Im}(\lambda_{311}\lambda^*_{322}) \ ,\nonumber\\
x_2&=& {\rm Im}(\lambda_{211}\lambda^*_{233}) \ ,\nonumber\\
x_3&=& \sum_{i=2,3} {\rm Im}(\lambda_{i11}\lambda'^*_{i11}) \ ,\nonumber \\
x_4&=& \sum_{i=2,3} {\rm Im}(\lambda_{i11}\lambda'^*_{i22}) \ ,\nonumber\\
x_5&=& \sum_{i=2,3} {\rm Im}(\lambda_{i11}\lambda'^*_{i33}) \ ,\nonumber \\
x_6&=& \sum_{i=1,3} {\rm Im}(\lambda_{i22}\lambda'^*_{i11}) \ ,\nonumber\\
x_7&=& \sum_{i=1,2} {\rm Im}(\lambda_{i33}\lambda'^*_{i11}) \ ,\nonumber\\
x_8&=& \sum_{i=1,2,3} {\rm Im}(\lambda'_{i11}\lambda'^*_{i22}) \ ,\nonumber\\
x_9&=& \sum_{i=1,2,3} {\rm Im}(\lambda'_{i11}\lambda'^*_{i33}) \ ,\nonumber\\
x_{10}&=& \sum_{i=1,2,3} {\rm Im}(\lambda'_{i22}\lambda'^*_{i33}) \ .
\label{eq:x1tox10}
\end{eqnarray}
These variables are the minimal set of variables linearly independent in the EDM relations (to see the relation, see Ref. \cite{yamanaka1}).
According to the classification of RPV bilinears, $x_1$ and $x_2$ belongs to 
the type 1.
Variables $x_3$, $x_4$ and $x_5$ are of the type 2, $x_6$ and $x_7$ of the type 3 and finally $x_8$, $x_9$ and $x_{10}$ of the type 4.
We must note that this decomposition has been made possible since we have assumed that the sneutrino masses (and charged slepton also) are degenerate.
If this would not be the case, the combinations of RPV couplings (\ref{eq:x1tox10}) would be much more involved.

\subsection{Phenomenological constraints on RPV couplings}

The above set of RPV bilinears are bounded in absolute values from other experimental data.
As we are considering the full RPV parameter space which can be constrained by EDM experiments, these absolute upper limits provided 
by other experiments should also be taken into account.
They are essentially derived from the fundamental precision test in experiments and are given in Table \ref{table:rpvlimit}.
\begin{table}[h]
\begin{center}
\begin{tabular}{lll}
\hline \hline
RPV coupling&Upper limit & Source\\
\hline
$|\lambda_{211}|$, $|\lambda_{122}|$&$0.05 [m_{\tilde e_R}]$& Universality \cite{kao}\\
$|\lambda_{311}|$, $|\lambda_{133}|$&$0.03 [m_{\tilde e_R}]$& Universality \cite{kao}\\
$|\lambda_{322}|$, $|\lambda_{233}|$&$0.05 [m_{\tilde e_R}]$& Universality \cite{kao}\\
$|\lambda'_{i11}|$, $|\lambda'_{i22}|_{(i=1,2,3)}$&$5.7 \times 10^{-3} [m_{\tilde d_R}]$&$K \rightarrow \pi \nu \bar \nu$ \cite{rpvkdecay,expkdecay,smkdecay}\\
$|\lambda'_{233}|$&$0.14 [m_{\tilde d_R}]$&$K \rightarrow \pi \nu \bar \nu$ \cite{rpvkdecay,expkdecay,smkdecay}\\
$|\lambda'_{311}|$&$10^{-3}[m_{\tilde \nu_\tau}]$ &Collider \cite{rpvcollider}\\
$|\lambda'_{111}|$&$2.8\times 10^{-3}[m_{\rm SUSY }]^{\frac{3}{2}} $&$0\nu \beta \beta$ \cite{rpvdoublebeta}\\
$|\lambda'_{133}|$&$3.5 \times 10^{-3}$&$\nu$ mass \cite{sallydawson}\\
$|\lambda'_{333}|$&$0.12[m_{\tilde d_R}]$&$B$ decay \cite{rpvbdecay}\\
\hline
\end{tabular}
\end{center}
\caption{Limits to RPV coupling from other experiments, given in absolute values \cite{rpvphenomenology}. 
The notation $[m]$ in the column of "Upper limit" denotes the mass $m$ in unit of 100 GeV.
}
\label{table:rpvlimit}
\end{table}
These limits are quite useful because the multidimensional region of the RPV coupling parameter space is now bounded by them.
Thanks to these limits, we obtain the 20 linear inequalities which corresponds to a finite ``box" in 10 dimensional RPV parameter space as follows:
\begin{equation}
\begin{array}{cccccc}
-0.15 & <& x_1& < & 0.15& ,\cr
-0.25 & <& x_2& <  &0.25 & ,\cr
-3.2 \times 10^{-2} & <& x_3& <  &3.2 \times 10^{-2} & ,\cr
-4.6 \times 10^{-2} & <& x_4& <  &4.6 \times 10^{-2} & ,\cr
-1.1 & <& x_5& <  &1.1 & ,\cr
-5.0 \times 10^{-2} & <& x_6& <  &5.0 \times 10^{-2} & ,\cr
-5.5 \times 10^{-2} & <& x_7& <  &5.5 \times 10^{-2} & ,\cr
-8.9\times 10^{-3} & <& x_8& < & 8.9\times 10^{-3} &,\cr
-9.5 \times 10^{-2} & <& x_9& <  &9.5 \times 10^{-2} & , \cr
-0.15 & <& x_{10}& <  &0.15 & ,\cr
\end{array}
\label{eq:rpvboxlimit}
\end {equation}
All in (\ref{eq:rpvboxlimit}) are given by assuming the SUSY masses to be equal 1 TeV.
In particular, sneutrino masses are assumed to degenerate.

The P, CP-odd observables can be expressed, to the leading order in RPV bilinears, as
\begin{equation}
d_a = \sum_{i=1}^{10}  c_{ai} x_i \ ,
\label{eq:sum}
\end{equation}
where $a$ is the label of the system (for example, He for $^3$He nucleus).
$c_{ai}$ are coefficients of the linear dependences of the P, CP-odd observable $a$ on the RPV bilinears.
The EDM-constraints are then expressed in the form of inequalities
\begin{equation}
- d_a^{\rm exp} < \sum_i c_{ai} x_i < d_a^{\rm exp} \ ,
\label{eq:rpvedmlimit}
\end{equation}
for $a=$Xe, Hg, Tl, YbF, ThO and $n$.
$d_a^{\rm exp}$ is the current experimental upper limits of the corresponding EDM observables:
\begin{eqnarray}
d_{\rm Xe}^{\rm exp} &=& 4.0 \times 10^{-27} e \, {\rm cm} \ ,
\nonumber\\
d_{\rm Tl}^{\rm exp} &=& 9 \times 10^{-25} e \, {\rm cm} \ , 
\nonumber\\
d_n^{\rm exp} &=& 2.9 \times 10^{-26} e \, {\rm cm} \ , 
\nonumber \\
d_{\rm Hg}^{\rm exp} &=& 3.1 \times 10^{-29} e \, {\rm cm} \ , 
\nonumber\\
d_{\rm YbF}^{\rm exp} &=& K_{\rm YbF} \times 1.05 \times 10^{-27} e \, {\rm cm} \ , \nonumber\\
d_{\rm ThO}^{\rm exp} &=& K_{\rm ThO} \times 8.7 \times 10^{-29} e \, {\rm cm} \ ,
\label{eq:currentexpupperlimit}
\end{eqnarray}
{\colourcyan{where $K_{\rm YbF}$ and $K_{\rm ThO}$ are the electron EDM enhancement factors for which explicit values are not needed in this work.}}

The computational strategy of our problem is then to find the maximally possible values of EDMs $d_a$ ($a$=$p$, $D$, He, Rn, Ra, Fr) together with $R$ ($R$-correlation in neutron beta decay) in the parameter space $(x_{1}, \cdots , x_{10})$ linearly constrained by Eqs. (\ref{eq:rpvboxlimit}) and (\ref{eq:rpvedmlimit}).

\subsection{Linear coefficients of CP-odd observables}

The final step of the setup is to set the inequalities due to the EDM-constraints from $^{129}$Xe, $^{205}$Tl, $^{199}$Hg atoms, YbF, ThO molecules, and neutron, and relations of the P, CP-odd observables we want to maximize (EDMs of the proton, deuteron, $^3$He nucleus, $^{211}$Rn, $^{225}$Ra, $^{210}$Fr atoms, muon, and the $R$-correlation of the neutron beta decay).
In this part, the scalar content of nucleon is abreviated as $\langle \bar qq \rangle \equiv \langle p | \bar qq | p \rangle$.

They are given as follows.

\subsubsection*{Paramagnetic atoms, molecules}

Here we present the linear coefficients of the EDM of the paramagnetic atoms and molecules ($^{205}$Tl, $^{210}$Fr, YbF and ThO).
\begin{eqnarray}
c_{a1} 
&=&
-K_a d_{\rm BZ} (2,1, -1,-1 )
\, ,
\nonumber\\
c_{a2} 
&=&
-K_a d_{\rm BZ} (3,1 , -1,-1 )
\, ,
\nonumber\\
c_{a3} 
&=&
K_a 
R^{\rm SP}_a \Biggl(
\frac{Z}{A} \langle \bar dd \rangle
+\frac{N}{A} \langle \bar uu \rangle
\Biggr)
\frac{1}{\sqrt{2} m_{\tilde \nu_i}^2 G_F}
 ,
\nonumber\\
c_{a4} 
&=&
-K_a d_{\rm BZ} (2,1 , -1/3,-1 )
\nonumber\\
&&
+K_a 
R^{\rm SP}_a 
\langle \bar ss \rangle
\frac{1}{\sqrt{2} m_{\tilde \nu_i}^2 G_F}
,
\nonumber\\
c_{a5} 
&=&
-K_a d_{\rm BZ} (3,1 , -1/3,-1 )
\nonumber\\
&&
+K_a 
R^{\rm SP}_a 
\langle \bar bb \rangle
\frac{1}{\sqrt{2} m_{\tilde \nu_i}^2 G_F}
,
\label{eq:paramagneticcoef}
\end{eqnarray}
where $a={\rm Tl},{\rm Fr}, {\rm YbF}, {\rm ThO}$.
The atomic level parameters are given in Table \ref{table:paramagnetic}.

\begin{table}
\caption{
The electron EDM enhancement factors $K_a$ and the relative linear coefficients of the scalar-type P, CP-odd e-N interactions $R^{\rm SP}_a $ for the $^{205}$Tl \cite{flambaumtl}, $^{210}$Fr \cite{flambaumfr,kelly,mukherjee}, YbF \cite{flambaumybftho}, and ThO \cite{flambaumybftho} systems.
The enhancement factors for the molecular systems are not shown, since the experimental constraints are given in terms of the electron EDM.
We also show the proton and neutron numbers $Z$ and $N$ of the nuclei.
}
\begin{ruledtabular}
\begin{tabular}{lcccc}
  & $K_a$ & $R^{\rm SP}_a $ &$Z$ & $N$\\ 
\hline
$^{205}$Tl & $-$582 & $1.2 \times 10^{-20}e$ cm & 81 & 124 \\
$^{210}$Fr & 910 & $1.2 \times 10^{-20}e$ cm & 87 & 123 \\
YbF & $-$ & $8.8 \times 10^{-21}e$ cm & 70 & 104 \\
ThO & $-$ & $1.3 \times 10^{-20}e$ cm & 90 & 142 \\
\end{tabular}
\end{ruledtabular}

\label{table:paramagnetic}
\end{table}

\subsubsection*{Diamagnetic atoms}

The linear coefficients for the diamagnetic atoms on the other hand are given as follows:
\begin{eqnarray}
c_{a1} 
&=&
-K_a^{d_e} d_{\rm BZ} (2,1 , -1,-1 )
\, ,
\nonumber\\
c_{a2} 
&=&
-K_a^{d_e} d_{\rm BZ} (3,1 , -1,-1 )
\, ,
\nonumber\\
c_{a3} 
&=&
\frac{1}{\sqrt{2} m_{\tilde \nu_i}^2 G_F}
\Biggl\{
K^{\rm SP}_a \Biggl(
\frac{Z}{A} \langle \bar dd \rangle
+\frac{N}{A} \langle \bar uu \rangle
\Biggr)
\nonumber\\
&&\hspace{5.5em}
- K^{\rm PS}_a \Bigl[
\langle \sigma_p^a \rangle \langle \bar di\gamma_5 d \rangle
+ \langle \sigma_n^a \rangle \langle \bar ui\gamma_5 u \rangle
\Bigr]
\Biggr\}
,
\nonumber
\\
c_{a4} 
&=&
-K_a^{d_e} d_{\rm BZ} (2,1 , -1/3,-1 )
\nonumber\\
&&
+\Bigl[
K^{\rm SP}_a 
\langle \bar ss \rangle
- K^{\rm PS}_a 
\langle \sigma_N^a \rangle \langle \bar si\gamma_5 s \rangle
\Bigr]
\frac{1}{\sqrt{2} m_{\tilde \nu_i}^2 G_F}
\, ,
\nonumber\\
c_{a5} 
&=&
-K_a^{d_e} d_{\rm BZ} (3,1 , -1/3,-1 )
\nonumber\\
&&
+\Bigl[
K^{\rm SP}_a 
\langle \bar bb \rangle
- K^{\rm PS}_a 
\langle \sigma_N^a \rangle
 \langle \bar b i\gamma_5 b \rangle
\Bigr]
\frac{1}{\sqrt{2} m_{\tilde \nu_i}^2 G_F}
,
\nonumber\\
c_{a6} 
&=&
K^{S}_a 
 \Bigl[
s_p^a \delta d
+ s_n^a \delta u
\Bigr]
d_{\rm BZ} (2,1 , -1,-1/3 )
,
\nonumber\\
c_{a7} 
&=&
K^{S}_a 
 \Bigl[
s_p^a \delta d
+ s_n^a \delta u
\Bigr]
d_{\rm BZ} (3,1 , -1,-1/3 )
,
\nonumber\\
c_{a8} 
&=&
-K^{S}_a 
\Biggl\{
 \Bigl[
s_p^a 
\delta d
+ 
s_n^a \delta u
\Bigr]
d_{\rm BZ} (2,1 , -1/3,-1/3 )
\nonumber\\
&& \hspace{3em}
+ e \Bigl[
s_p^a \tilde \rho^d_p
+ 
s_n^a \tilde \rho^d_n
\Bigr]
d_{\rm BZ}^c (2,1)
\nonumber\\
&& \hspace{3em}
+\left[ a_a^{(0)} \tilde \omega_{(0)} -a_a^{(1)} \tilde \omega_{(1)} \right] g_{\pi NN} e\, d_{\rm BZ}^c (2,1 )
\nonumber\\
&& \hspace{3em}
-a_a^{(1)} g_{\pi NN} \frac{e f_\pi m_\pi^2 \langle \bar s s \rangle}{2 m_d} \frac{1}{2 m_{\tilde \nu_i}^2 }
\Biggr\}
,
\nonumber
\end{eqnarray}
\begin{eqnarray}
c_{a9} 
&=&
-K^{S}_a 
\Biggl\{
 \Bigl[
s_p^a \delta d
+ 
s_n^a \delta u
\Bigr]
d_{\rm BZ} (3,1 , -1/3,-1/3 )
\nonumber\\
&& \hspace{3em}
+ e \Bigl[
s_p^a \tilde \rho^d_p
+ 
s_n^a \tilde \rho^d_n
\Bigr]
d_{\rm BZ}^c (3,1 )
\nonumber\\
&& \hspace{3em}
+\left[ a_a^{(0)} \tilde \omega_{(0)} -a_a^{(1)} \tilde \omega_{(1)} \right] g_{\pi NN} e\, d_{\rm BZ}^c (3,1 )
\nonumber\\
&& \hspace{3em}
-a_a^{(1)} g_{\pi NN} \frac{e f_\pi m_\pi^2 \langle \bar bb \rangle}{2 m_d} \frac{1}{2 m_{\tilde \nu_i}^2 }
\Biggr\}
,
\nonumber\\
c_{a10} 
&=&
-K^{S}_a 
\Biggl\{
(s_p^a +s_n^a ) \delta s\:
d_{\rm BZ} (3,2 , -1/3,-1/3 )
\nonumber\\
&& \hspace{3em}
+ e \Bigl[
s_p^a \tilde \rho^s_p
+ 
s_n^a \tilde \rho^s_n
\Bigr]
d_{\rm BZ}^c (3,2 )
\Biggr\}
,
\label{eq:diamagnetic}
\end{eqnarray}
where $a={\rm Xe},{\rm Hg}, {\rm Rn}, {\rm Ra}$.
The atomic level parameters are given in Tables \ref{table:diamagnetic} and \ref{table:diamagnetic2}.
The nuclear level parameters for the ${\rm Hg}$, ${\rm Rn}$, and ${\rm Ra}$ nuclei are given in Table \ref{table:schiffmoment}.
For the nuclear level input parameters of the $^{129}$Xe atom, we have used the result of the shell model analysis where the nucleon spin coefficients are given by $s_p^{\rm Xe} = +0.0061\, {\rm fm}^{2}$ and $s_n^{\rm Xe} = -0.3169\, {\rm fm}^{2}$ \cite{yoshinaga1}, and the dependence of the $^{129}$Xe nuclear Schiff moment on the P, CP-odd pion-nucleon interactions is given by $a_{\rm Xe}^{(0)} = -5.07 \times 10^{-4} e\, {\rm fm}^3$ and $a_{\rm Xe}^{(1)} =-3.99 \times 10^{-4} e\, {\rm fm}^3$ \cite{yoshinaga2}.

\begin{table*}
\caption{
The atomic level parameters for diamagnetic atoms ($^{129}$Xe, $^{199}$Hg, $^{211}$Rn, and $^{225}$Ra) \cite{flambaumdiamagnetic}.
The effects of the tensor- ($C_N^{\rm T}$), pseudoscalar-type P, CP-odd interactions ($C_N^{\rm PS}$), and the nuclear Schiff moment to the EDMs of the diamagnetic atoms are given.
We also show the proton and neutron numbers $Z$ and $N$ of the nuclei, and the nuclear magnetic moment in unit of the nuclear magneton.
}
\begin{ruledtabular}
\begin{tabular}{lccccccc}
  &$Z$ & $N$&$\langle \sigma_N^a \rangle$ & $\mu_a$ & $K^{\rm T}_a $ ($\langle \sigma^a_N \rangle  e$ cm) & $K^{\rm PS}_a $ ($\langle \sigma^a_N \rangle  e$ cm) &$K^S_a $ (cm/fm$^3$) \\ 
\hline
$^{129}$Xe & 54 & 75 & 1 & -0.7778 & $5.7 \times 10^{-21} $ & $1.6 \times 10^{-23} $ & $3.8 \times 10^{-18} $  \\
$^{199}$Hg & 80 & 119 & $-\frac{1}{3}$ &0.5059 & $-5.1 \times 10^{-20}$ & $-1.8 \times 10^{-22}$ & $- 2.6 \times 10^{-17} $ \\
$^{211}$Rn & 86 & 125 & $-\frac{1}{3}$ & 0.601 & $5.6 \times 10^{-20}$ & $2.1 \times 10^{-22}$ & $3.3 \times 10^{-17} $ \\
$^{225}$Ra & 88 & 137 & 1 & -0.734 & $-1.8 \times 10^{-19}$  & $-6.4 \times 10^{-22}$  & $- 8.8 \times 10^{-17} $ \\
\end{tabular}
\end{ruledtabular}

\label{table:diamagnetic}
\end{table*}

\begin{table}
\caption{
The atomic level parameters for diamagnetic atoms ($^{129}$Xe, $^{199}$Hg, $^{211}$Rn, and $^{225}$Ra) calculated using the analytic formulae \cite{flambaumformula}.
The electron EDM enhancement factors and the linear coefficients of the scalar-type P, CP-odd e-N interactions ($C^{\rm SP}_N $) are shown.
}
\begin{ruledtabular}
\begin{tabular}{lcc}
 & $K_a^{d_e}$ & $K^{\rm SP}_a $ ($e$ cm)  \\ 
\hline
$^{129}$Xe & $-9.8 \times 10^{-4}$ & $-6.2 \times 10^{-23} $   \\
$^{199}$Hg  & $-7.9 \times 10^{-3}$ & $-5.1 \times 10^{-22}$ \\
$^{211}$Rn & $1.1 \times 10^{-2}$ & $7.0 \times 10^{-22}$  \\
$^{225}$Ra  & $4.3 \times 10^{-2}$ & $2.9 \times 10^{-21}$ \\
\end{tabular}
\end{ruledtabular}

\label{table:diamagnetic2}
\end{table}

\subsubsection*{Nucleon EDM}

The linear coefficients for the neutron EDM are given by
\begin{eqnarray}
c_{n6} 
&=&
\delta u\: d_{\rm BZ} (2,1 , -1,-1/3 )
,
\nonumber\\
c_{n7} 
&=&
\delta u\: d_{\rm BZ} (3,1 , -1,-1/3 )
,
\nonumber\\
c_{n8} 
&=&
-\delta u\: d_{\rm BZ} (2,1 , -1/3,-1/3 )
- e  \tilde \rho^d_n d_{\rm BZ}^c (2,1 )
,
\nonumber\\
c_{n9} 
&=&
-\delta u\: d_{\rm BZ} (3,1 , -1/3,-1/3 )
- e  \tilde \rho^d_n d_{\rm BZ}^c (3,1 )
,
\nonumber\\
c_{n10} &=&
-\delta s\: d_{\rm BZ} (3,2 , -1/3,-1/3 )
- e \tilde \rho^s_n d_{\rm BZ}^c (3,2 )
. \ \ \ \ 
\label{eq:ncoef}
\end{eqnarray}
For the proton EDM, the coefficients are obtained just by the replacement $\delta u \rightarrow \delta d$, $\tilde \rho^d_n \rightarrow \tilde \rho^d_p$, and  $\tilde \rho^s_n \rightarrow \tilde \rho^s_p$ in (\ref{eq:ncoef}).

\subsubsection*{Light nuclei}

The linear coefficients for the light nuclear EDM are
\begin{eqnarray}
c_{a6} 
&=&
\Bigl[ \langle \sigma_n^a \rangle \delta u +\langle \sigma_p^a \rangle \delta d \Bigr] d_{\rm BZ} (2,1 , -1,-1/3 )
,
\nonumber\\
c_{a7} 
&=&
\Bigl[ \langle \sigma_n^a \rangle \delta u +\langle \sigma_p^a \rangle \delta d \Bigr] d_{\rm BZ} (3,1 , -1,-1/3 )
,
\nonumber
\end{eqnarray}
\begin{eqnarray}
c_{a8}
&=&
-\Bigl[ \langle \sigma_n^a \rangle \delta u +\langle \sigma_p^a \rangle \delta d \Bigr] d_{\rm BZ} (2,1 , -1/3,-1/3 )
\nonumber\\
&&
-\Bigl[ \langle \sigma_n^a \rangle \tilde \rho^d_n +\langle \sigma_p^a \rangle \tilde \rho^d_p \Bigr] e d_{\rm BZ}^c (2,1 )
\nonumber\\
&&
-K_a^{(0)} g_{\pi NN} e
\tilde \omega_{(0)}  d_{\rm BZ}^c (2,1 )
\nonumber\\
&&
+K_a^{(1)} g_{\pi NN} e
\Biggl\{
\tilde \omega_{(1)}  d_{\rm BZ}^c (2,1 )
+  \frac{ f_\pi m_\pi^2 \langle \bar ss \rangle}{2 m_d\cdot 2 m_{\tilde \nu_i}^2} 
\Biggr\}
,
\nonumber\\
c_{a9}
&=&
-\Bigl[ \langle \sigma_n^a \rangle \delta u +\langle \sigma_p^a \rangle \delta d \Bigr] d_{\rm BZ} (3,1, -1/3,-1/3 )
\nonumber\\
&&
-\Bigl[ \langle \sigma_n^a \rangle \tilde \rho^d_n +\langle \sigma_p^a \rangle \tilde \rho^d_p \Bigr] e d_{\rm BZ}^c (3,1 )
\nonumber\\
&&
-K_a^{(0)} g_{\pi NN} e
\tilde \omega_{(0)}  d_{\rm BZ}^c (3,1 )
\nonumber\\
&&
+K_a^{(1)} g_{\pi NN} e
\Biggl\{
\tilde \omega_{(1)}  d_{\rm BZ}^c (3,1 )
+  \frac{ f_\pi m_\pi^2 \langle \bar bb \rangle}{2 m_d\cdot 2 m_{\tilde \nu_i}^2} 
\Biggr\}
,
\nonumber\\
c_{a10}
&=&
-\Bigl[ \langle \sigma_n^a \rangle +\langle \sigma_p^a \rangle \Bigl] \delta s\: d_{\rm BZ} (3,2 , -1/3,-1/3 )
\nonumber\\
&&
-\Bigl[ \langle \sigma_n^a \rangle \tilde \rho^s_n +\langle \sigma_p^a \rangle \tilde \rho^s_p \Bigl] e d_{\rm BZ}^c (3,2 )
,
\label{eq:light_nuclei}
\end{eqnarray}
where $a = d\, ,\, ^3{\rm He}$.
The nuclear level parameters are given in Table \ref{table:nuclearedm}.

\begin{table}
\caption{
Coefficients $a^{(i)}$ of the dependence of the nuclear EDM on the P, CP-odd pion-nucleon couplings ($d_a = g_{\pi NN} (a_a^{(0)} \bar g^{(0)}_{\pi NN} +a_a^{(1)} \bar g^{(1)}_{\pi NN} + a_a^{(2)} \bar g^{(2)}_{\pi NN})$).
The coefficients $a_a^{(i)}$ are expressed in unit of $10^{-13} e$ cm for the $d$ \cite{liu} and $^3$He \cite{gudkov,stetcu} nuclei. 
}
\begin{ruledtabular}
\begin{tabular}{lccccc}
& $-a^{(0)}_a$\ \  &$-a^{(1)}_a$ \ \ &$a^{(2)}_a$\ \ &$\langle \sigma^a_p \rangle$ &$\langle \sigma^a_n \rangle$ \ \  \\ 
\hline
$d$ & 0 & $-$0.015 & 0 &1 & 1 \\
$^3$He & $-$0.008 & $-$0.012 & 0.007 & $-$0.04 & 0.90  \\
\end{tabular}
\end{ruledtabular}
\label{table:nuclearedm}
\end{table}

\subsubsection*{$R$-correlation}

Finally the $R$-correlation in the neutron beta decay is given simply by
\begin{eqnarray}
R=c_{R3}x_{3}, 
\label{eq:Rc3x3}
\end{eqnarray}
where
\begin{eqnarray}
c_{R3} 
&=&
\frac{\sqrt{2} g_A (\langle \bar uu \rangle -\langle \bar dd \rangle )}{V_{ud} G_F (1+g_A^2) m_{\tilde e_{Li}}^2}
,
\label{eq:rcoef}
\end{eqnarray}
and the ratio $g_{A}=G_{A}/G_{V}$  is given by \cite{ucna}
\begin{eqnarray}
g_A =1.27590\pm 0.00239^{+0.00331}_{-0.00377} .
\label{eq:gagvratio}
\end{eqnarray}
The derivation of (\ref{eq:Rc3x3}) and (\ref{eq:rcoef}) is relegated in Appendix \ref{sec:r-corr}.

The numerical values of coefficients $c_{ai}$ are tabulated in 
Table \ref{table:addedonoctober5}
and 
Table \ref{table:addedonoctober1}.

\begin{table*}[htb]
\begin{center}
\begin{tabular}{c|cccccc}
\hline \hline
Obs. &$d_{\rm Tl}$&$d_{\rm Hg}$ &$d_{n}$&$d_{\rm Xe}$&$d_{\rm YbF}$&$d_{\rm ThO}$
\\
\hline
$c_{a1}$&$-3.0\times 10^{-22}$&$-4.1 \times 10^{-27}$&$0$&$-5.1 \times 10^{-28}$&$5.2 \times 10^{-25}$&$5.2 \times 10^{-25}$
\\
$c_{a2}$&$-3.5 \times 10^{-21}$&$-4.7 \times 10^{-26}$&$0$&$-5.8 \times 10^{-27}$&$5.9 \times 10^{-24}$&$5.9 \times 10^{-24}$
\\
$c_{a3}$&$-2.6 \times 10^{-18}$&$-7.4 \times 10^{-22}$&$0$&$-1.2 \times 10^{-22}$&$3.5 \times 10^{-21}$&$5.1 \times 10^{-21}$
\\
$c_{a4}$&$-4.2 \times 10^{-20}$&$8.1 \times 10^{-24}$&$0$&$2.5 \times 10^{-24}$&$5.3 \times 10^{-23}$&$7.9 \times 10^{-23}$
\\
$c_{a5}$&$-7.0\times 10^{-21}$&$-1.4 \times 10^{-25}$&$0$&$1.8 \times 10^{-26}$&$9.6 \times 10^{-24}$&$1.3 \times 10^{-23}$
\\
$c_{a6}$&$0$&$-2.1 \times 10^{-29}$&$-1.6 \times 10^{-25}$&$2.0 \times 10^{-30}$&$0$&$0$
\\
$c_{a7}$&$0$&$-2.6 \times 10^{-28}$&$-2.0 \times 10^{-24}$&$2.4 \times 10^{-29}$&$0$&$0$
\\
$c_{a8}$&$0$&$-5.6 \times 10^{-26}$&$-4.0 \times 10^{-23}$&$1.8 \times 10^{-28}$&$0$&$0$
\\
$c_{a9}$&$0$&$-7.5 \times 10^{-25}$&$-8.9 \times 10^{-22}$&$8.3 \times 10^{-27}$&$0$&$0$
\\
$c_{a10}$&$0$&$-2.2 \times 10^{-26}$&$-1.8 \times 10^{-22}$&$2.2 \times 10^{-27}$&$0$&$0$
\\
\hline
\end{tabular}
\end{center}
\caption{Coefficients $c_{ai}$  ($i=1, \cdots ,10$) for the EDMs of $a ( = $ 
$^{205}$Tl, $^{199}$Hg, neutron, $^{129}$Xe, YbF and ThO). The sneutrino mass was taken to be 1TeV. The unit is in $e$ cm.}
\label{table:addedonoctober5}
\end{table*}

\begin{table*}[htb]
\begin{center}
\begin{tabular}{c|ccccccc}
\hline \hline
 &$d_p$&$d_d$&$d_{\rm He}$ &$d_{\rm Rn}$&$d_{\rm Ra}$&$d_{\rm Fr}$&$R$\\
\hline
$c_{a1}$ & $0$  & $0$ & $0$  & $5.6 \times 10^{-27}$ & $2.2 \times 10^{-26}$ & $4.8 \times 10^{-22}$ & $0$ \\
$c_{a2}$ & $0$  & $0$ & $0$  & $6.4 \times 10^{-26}$ & $2.5 \times 10^{-25}$ & $5.4 \times 10^{-21}$ & $0$ \\
$c_{a3}$ & $0$  & $0$ & $0$  & $-4.4 \times 10^{-22}$ & $5.1 \times 10^{-21}$ & $4.3 \times 10^{-18}$ & $-1.3 \times 10^{-2}$ \\
$c_{a4}$ & $0$  & $0$ & $0$  & $1.7 \times 10^{-23}$ & $-1.1 \times 10^{-22}$ & $6.6 \times 10^{-20}$ & $0$ \\
$c_{a5}$ & $0$  & $0$ & $0$  & $8.4 \times 10^{-25}$ & $-5.8 \times 10^{-25}$ & $1.1 \times 10^{-20}$ & $0$ \\
$c_{a6}$ & $4.0 \times 10^{-26}$  & $-1.2 \times 10^{-25}$ & $-1.5 \times 10^{-25}$  & $8.1 \times 10^{-29}$ & $0$ & $0$ & $0$ \\
$c_{a7}$ & $4.9 \times 10^{-25}$  & $-1.5 \times 10^{-25}$ & $-1.8 \times 10^{-24}$  & $9.8 \times 10^{-28}$ & $0$ & $0$ & $0$ \\
$c_{a8}$ & $4.0 \times 10^{-23}$  & $-3.5 \times 10^{-22}$ & $-2.9 \times 10^{-22}$  & $1.7 \times 10^{-25}$ & $1.3 \times 10^{-22}$ & $0$ & $0$ \\
$c_{a9}$ & $8.9 \times 10^{-22}$  & $-3.6 \times 10^{-21}$ & $-3.2 \times 10^{-21}$  & $2.3 \times 10^{-24}$ & $1.3 \times 10^{-21}$ & $0$ & $0$ \\
$c_{a10}$ & $1.8 \times 10^{-22}$  & $-1.7 \times 10^{-25}$ & $-1.7 \times 10^{-22}$  & $8.9 \times 10^{-26}$ & $0$ & $0$ & $0$ \\
\hline
\end{tabular}
\end{center}
\caption{Coefficients $c_{ai}$ ($i=1, \cdots ,10$) for the EDMs of $a ( =$
proton, deuteron, $^{3}$He nucleus, $^{211}$Rn, $^{225}$Ra atoms,  Fr and the $R$-correlation of the neutron beta decay).
The sneutrino mass was taken to be 1TeV. The unit is in $e$ cm, except for the 
$R$-correlation which is dimensionless.
}
\label{table:addedonoctober1}
\end{table*}

\section{\label{sec:analysis}Results and analysis}

\subsection{Constraints to RPV couplings}

After performing the analysis using the linear programming method, we have found the upper limits on bilinears of RPV couplings listed in Table \ref{table:rpvlpresult1}.
We see that the RPV bilinears $x_2[= {\rm Im}(\lambda_{211}\lambda^*_{233})] $, $x_3[= \sum_{i=2,3} {\rm Im}(\lambda_{i11}\lambda'^*_{i11}) ]$, $x_4[= \sum_{i=2,3} {\rm Im}(\lambda_{i11}\lambda'^*_{i22})]$, $x_5[= \sum_{i=2,3} {\rm Im}(\lambda_{i11}\lambda'^*_{i33})]$, $x_9[= \sum_{i=1,2,3} {\rm Im}(\lambda'_{i11}\lambda'^*_{i33}) ]$ and $x_{10}[= \sum_{i=1,2,3} {\rm Im}(\lambda'_{i22}\lambda'^*_{i33}) ]$ are constrained in such a way that the limits obtained are apparently looser than those obtained in the previous analysis based on the dominance of a single RPV bilinear (see Table \ref{table:single_coupling}).
This illustrates the large degree of freedom of the RPV supersymmetric models which accommodate much larger regions of the couplings while keeping consistency with tight EDM-constraints.
Note that this analysis gives in some cases tighter upper bounds than the limits listed in Eq. (\ref{eq:rpvboxlimit}) on the basis of  other experimental sources in  Table \ref{table:rpvlimit}. 
We can safely say that the analysis leading to (\ref{eq:rpvboxlimit}) gives cruder upper bounds on the RPV couplings.

\begin{table*}[htb]
\begin{center}
\begin{tabular}{l|ccc}
\hline \hline
RPV bilinears &Limits ($m_{\rm SUSY}=1\, {\rm TeV}$)&Limits ($m_{\rm SUSY}=5\, {\rm TeV}$) &Limits ($m_{\rm SUSY}=10\, {\rm TeV}$) \\
\hline
$|x_1|(= |{\rm Im}(\lambda_{311}\lambda^*_{322})|) $& 0.15 & 3.75 & 15\\
$|x_2|(= |{\rm Im}(\lambda_{211}\lambda^*_{233})|) $& 0.12 & 3.0 & 12\\
$|x_3|(= |\sum_{i=2,3} {\rm Im}(\lambda_{i11}\lambda'^*_{i11})| )$&$1.8\times 10^{-4}$& $5.5\times 10^{-3}$ &$2.4\times 10^{-2}$\\
$|x_4|(= |\sum_{i=2,3} {\rm Im}(\lambda_{i11}\lambda'^*_{i22})| )$&$1.1\times 10^{-2}$& 0.33 & 1.4\\
$|x_5|(= |\sum_{i=2,3} {\rm Im}(\lambda_{i11}\lambda'^*_{i33})| )$&0.19& 5.8 &25\\
$|x_6|(= |\sum_{i=1,3} {\rm Im}(\lambda_{i22}\lambda'^*_{i11})| )$&$5.0\times 10^{-2}$& 2.6 & 14 \\
$|x_7|(=|\sum_{i=1,2} {\rm Im}(\lambda_{i33}\lambda'^*_{i11})| )$&$5.5\times 10^{-2}$& 2.2 & 11 \\
$|x_8|(= |\sum_{i=1,2,3} {\rm Im}(\lambda'_{i11}\lambda'^*_{i22})|)$&$8.9 \times 10^{-3}$& 0.38 & 2.0 \\
$|x_9|(=|\sum_{i=1,2,3} {\rm Im}(\lambda'_{i11}\lambda'^*_{i33})| )$&$3.0 \times 10^{-2}$& 0.76 & 3.1 \\
$|x_{10}|(= |\sum_{i=1,2,3} {\rm Im}(\lambda'_{i22}\lambda'^*_{i33})| )$&0.15& 3.7 & 15 \\
\hline
\end{tabular}
\end{center}
\caption{
Upper limits on the absolute value of bilinears of RPV couplings found by linear programming analysis.
}
\label{table:rpvlpresult1}
\end{table*}

\begin{table*}[htb]
\begin{center}
\begin{tabular}{l|ccc}
\hline \hline
RPV bilinears &Limits ($m_{\rm SUSY}=1\, {\rm TeV}$)&Limits ($m_{\rm SUSY}=5\, {\rm TeV}$) &Limits ($m_{\rm SUSY}=10\, {\rm TeV}$) \\
\hline
$|x_1|(= |{\rm Im}(\lambda_{311}\lambda^*_{322})|) $& $1.7 \times 10^{-4}$ & $3.4 \times 10^{-3}$ & $1.2 \times 10^{-2}$ \\
$|x_2|(= |{\rm Im}(\lambda_{211}\lambda^*_{233})|) $& $1.5 \times 10^{-5}$ & $2.7 \times 10^{-4}$ & $9.7 \times 10^{-4}$ \\
$|x_3|(= |\sum_{i=2,3} {\rm Im}(\lambda_{i11}\lambda'^*_{i11})| )$& $1.7\times 10^{-8}$& $4.2\times 10^{-7}$ &$1.7\times 10^{-6}$\\
$|x_4|(= |\sum_{i=2,3} {\rm Im}(\lambda_{i11}\lambda'^*_{i22})| )$&$1.1\times 10^{-6}$ & $2.8 \times 10^{-5}$ & $1.1 \times 10^{-4}$ \\
$|x_5|(= |\sum_{i=2,3} {\rm Im}(\lambda_{i11}\lambda'^*_{i33})| )$&$6.9\times 10^{-6}$ & $1.6 \times 10^{-4}$ &$6.5 \times 10^{-4}$\\
$|x_6|(= |\sum_{i=1,3} {\rm Im}(\lambda_{i22}\lambda'^*_{i11})| )$&0.18 & 3.5 & 13 \\
$|x_7|(=|\sum_{i=1,2} {\rm Im}(\lambda_{i33}\lambda'^*_{i11})| )$&$1.5\times 10^{-2}$& 0.26 & 0.93 \\
$|x_8|(= |\sum_{i=1,2,3} {\rm Im}(\lambda'_{i11}\lambda'^*_{i22})|)$&$7.3 \times 10^{-4}$& $1.5 \times 10^{-2}$ & $5.7 \times 10^{-2}$ \\
$|x_9|(=|\sum_{i=1,2,3} {\rm Im}(\lambda'_{i11}\lambda'^*_{i33})| )$&$3.2 \times 10^{-5}$& $5.9 \times 10^{-4}$ & $2.1 \times 10^{-3}$ \\
$|x_{10}|(= |\sum_{i=1,2,3} {\rm Im}(\lambda'_{i22}\lambda'^*_{i33})| )$&$1.6\times 10^{-4}$ & $3.0 \times 10^{-3}$ & $1.1 \times 10^{-2}$ \\
\hline
\end{tabular}
\end{center}
\caption{
Upper limits on the absolute value of bilinears of RPV couplings obtained with the assumption of the single coupling dominance.
The RPV bilinears $x_1$, $x_2$, $x_3$, $x_4$, and $x_5$ are constrained by the experimental data of the ThO molecule \cite{acme}.
The RPV bilinears $x_6$, $x_7$, $x_8$, $x_9$, and $x_{10}$ are constrained by the experimental data of the neutron EDM \cite{baker}.
}
\label{table:single_coupling}
\end{table*}

\subsection{Maximal prediction of the EDMs of the proton, deuteron, $^3$He nucleus, $^{ 211}$Rn, $^{225}$Ra, $^{ 210}$Fr atoms, and the $R$-correlation}

Next, we have made a prediction of the maximal value of the EDMs of the proton, deuteron, $^3$He nucleus, $^{211}$Rn, $^{225}$Ra, $^{ 210}$Fr atoms, and the $R$-correlation within the linear programming method.
This can be achieved by maximizing the linear relations constructed out of the coefficients of the above systems within the EDM-constraints of $^{205}$Tl, $^{199}$Hg, $^{129}$Xe, YbF, ThO, and neutron.

\begin{table*}[htb]
\begin{center}
\begin{tabular}{c|ccccccc}
\hline \hline
 &$d_p$&$d_d$&$d_{\rm He}$ &$d_{\rm Rn}$&$d_{\rm Ra}$&$d_{\rm Fr}$&$R$\\
\hline
Max. & $1.7\times 10^{-25}$  & $1.1\times 10^{-22}$ & $7.3 \times 10^{-23}$  &$ 9.5 \times 10^{-26}$ &$ 4.1 \times 10^{-23}$ &$ 
{\colourcyan{
3.1 \times 10^{-24}
}}$&$ 2.4\times 10^{-6}$ \\
\hline
$x_1$ & $-0.15$  & 0.15 & 0.15  & 0.15 & $-0.15$ & 0.15 & 0.15 \\ 
$x_2$ & $-9.2 \times 10^{-2}$ &$9.1 \times 10^{-2}$ & $2.2 \times 10^{-3}$  & $-0.11$ & $-9.3\times 10^{-2}$& $-0.12$ & $-0.11$ \\ 
$x_3$ & $-1.8 \times 10^{-4}$ &$1.8 \times 10^{-4}$ &$2.0 \times 10^{-5}$ &$-1.8 \times 10^{-4}$ &$-1.8 \times 10^{-4}$ & $-1.8 \times 10^{-4}$ & $-1.8\times 10^{-4}$\\ 
$x_4$ & $-1.1 \times 10^{-2}$ &$1.1 \times 10^{-2}$ &$-6.7 \times 10^{-4}$ &$-1.1 \times 10^{-2}$ & $-1.1 \times 10^{-2}$& $-1.1 \times 10^{-2}$ & $-1.1 \times 10^{-2}$ \\ 
$x_5$ & 0.19 & $-0.19$ & $-4.9 \times 10^{-3}$ & 0.19 & 0.19 & 0.19 & 0.19 \\ 
$x_6$ & $-5.0 \times 10^{-2}$ &$5.0 \times 10^{-2}$ & $5.0 \times 10^{-2}$ & $-5.0 \times 10^{-2}$ & $-5.0 \times 10^{-2}$& $-5.0 \times 10^{-2}$ & $-5.0 \times 10^{-2}$ \\ 
$x_7$ & $-5.5 \times 10^{-2}$ &$5.5 \times 10^{-2}$ & $5.5 \times 10^{-2}$ & $-5.5 \times 10^{-2}$  & $-5.5 \times 10^{-2}$  & $-5.5 \times 10^{-2}$ & $-5.5 \times 10^{-2}$ \\ 
$x_8$ & $-8.9 \times 10^{-3}$ &$-8.9 \times 10^{-3}$ & $-8.9 \times 10^{-3}$ & $8.9 \times 10^{-3}$ & $8.9 \times 10^{-3}$ & $8.9 \times 10^{-3}$ & $8.9 \times 10^{-3}$ \\ 
$x_9$ & $3.0 \times 10^{-2}$ & $-3.0 \times 10^{-2}$ & $-3.0 \times 10^{-2}$ & $3.0 \times 10^{-2}$ & $3.0 \times 10^{-2}$ & $3.0 \times 10^{-2}$ & $3.0 \times 10^{-2}$ \\ 
$x_{10}$ &  $-0.15$ & $0.15$ &  $0.15$ &  $-0.15$ &  $-0.15$ &  $-0.15$ & $-0.15$\\ 
\hline
\end{tabular}
\end{center}
\caption{
Maximal predictions of the EDMs of the proton, deuteron, $^3$He nucleus, $^{211}$Rn, $^{225}$Ra, and $^{210}$Fr atoms, and the $R$-correlation of the neutron beta decay, within the constraints of the $^{205}$Tl, $^{199}$Hg, $^{129}$Xe, YbF, ThO, and neutron EDM experiments.
Coordinates $x_i$ maximizing the observables are also shown.
The EDMs are expressed in unit of $e\, {\rm cm}$.
The sparticle mass $m_{\rm SUSY}$ has been taken to be 1 TeV.
}
\label{table:rpvlpresult}
\end{table*}
The obtained maximal predictions  are summarized in Table \ref{table:rpvlpresult} for $m_{\rm SUSY}=1\:{\rm TeV}$.
We also show the result in Table \ref{table:rpvlpresult5TeV} obtained exactly in the same way, but with $m_{\rm SUSY}=5$ TeV.
The predictions obtained here are on the same order of magnitude, since the constraints on biliears of RPV couplings given by EDMs and by other experiments (see Table \ref{table:rpvlimit}) have similar scaling in sparticle masses.
\begin{table*}[htb]
\begin{center}
\begin{tabular}{c|ccccccc}
\hline \hline
 &$d_p$&$d_d$&$d_{\rm He}$ &$d_{\rm Rn}$&$d_{\rm Ra}$&$d_{\rm Fr}$&$R$\\
\hline
Max. & $2.9\times 10^{-25}$  & $1.5\times 10^{-22}$ & $1.0 \times 10^{-22}$  &$ 1.3 \times 10^{-25}$ &$ 5.6 \times 10^{-23}$ &$ 
{\colourcyan{
3.7 \times 10^{-24}
}}
$ &$ 2.9\times 10^{-6}$ \\
\hline
\end{tabular}
\end{center}
\caption{
Maximal predictions of the EDMs of the proton, deuteron, $^3$He nucleus, $^{211}$Rn, $^{225}$Ra, and $^{210}$Fr  atoms, and the $R$-correlation of the neutron beta decay, within the constraints of the $^{205}$Tl, $^{199}$Hg, $^{129}$Xe, YbF, ThO and neutron EDM experiments for $m_{\rm SUSY} = 5$ TeV.
The EDMs are expressed in unit of $e\, {\rm cm}$ (the $R$-correlation is dimensionless).
}
\label{table:rpvlpresult5TeV}
\end{table*}

Let us compare our analysis with the previous ones relying on the assumption of the dominance of one single RPV bilinear summarized in Table \ref{table:single_coupling}.
Using these limits in  Table \ref{table:single_coupling}, we obtain the upper limits for P, CP-odd observables as in Table \ref{table:rpvsinglecouplingdominance}, which are going to be inspected in next generation experiments (see also Table \ref{table:rpvsinglecouplingdominance5TeV} for $m_{\rm SUSY}=5$ TeV).
\begin{table*}[htb]
\begin{center}
\begin{tabular}{c|cccccccc}
\hline \hline
&$d_{\rm Xe }$&$d_p$&$d_d$&$d_{\rm He}$ &$d_{\rm Rn}$&$d_{\rm Ra}$&$d_{\rm Fr}$&$R$\\
\hline
Max.&$2.8 \times 10^{-30}$ &$2.9\times 10^{-26}$&$2.6\times 10^{-25}$&$2.1\times 10^{-25}$&$1.2\times10^{-28}$&$9.3\times 10^{-26}$
&${\colourcyan{7.9\times 10^{-26}}}$& $2.3\times 10^{-10}$\\
\hline
\end{tabular}
\end{center}
\caption{Upper limits of the prepared experimental observables with the assumption of the single coupling dominance.
The unit is $e\, {\rm cm}$ for EDM observables (the $R$-correlation is dimensionless).
Sparticle masses $m_{\rm SUSY}$ has been taken to be 1 TeV.
}
\label{table:rpvsinglecouplingdominance}
\end{table*}
\begin{table*}[htb]
\begin{center}
\begin{tabular}{c|cccccccc}
\hline \hline
&$d_{\rm Xe }$&$d_p$&$d_d$&$d_{\rm He}$ &$d_{\rm Rn}$&$d_{\rm Ra}$&$d_{\rm Fr}$&$R$\\
\hline
Max. &$2.8\times 10^{-30}$ &$2.9\times 10^{-26}$&$2.3\times 10^{-25}$&$1.9\times 10^{-25}$&$1.1\times10^{-28}$&$8.4\times 10^{-26}$&$
{\colourcyan{7.9\times 10^{-26}}}$& $2.3\times 10^{-10}$\\
\hline
\end{tabular}
\end{center}
\caption{Upper limits of the prepared experimental observables with the assumption of the single coupling dominance for $m_{\rm SUSY}=5$ TeV.
The unit is $e\, {\rm cm}$ for EDM observables (the $R$-correlation is dimensionless).
}
\label{table:rpvsinglecouplingdominance5TeV}
\end{table*}
We see that all predictions with the assumption of the dominance of one single RPV bilinear are well below our maximal predictions using the linear programming method, by two to four orders of magnitudes.
This huge difference implies that choosing one of the couplings as the dominant one is not put on a sound basis and that there could occur conspiracy among several couplings  so that the EDM constraints are satisfied.
These configurations of RPV couplings have never been shed light in the previous analyses.
The linear programming method shows efficiency in finding such configurations for RPV parameters.
Below, we will try to explain in more detail the reasons for this large difference.

\subsubsection{
RPV couplings satisfying EDM-constraints of atoms and molecules}

As we have 
argued in Chapter \ref{sec:classification}, the atoms and molecules have a strong sensitivity to the type 2 ($x_3 , x_4$ and $x_5$) semi-leptonic RPV bilinears.
Let us see how the RPV bilinears arrange themselves and become large consistently with the EDM-constraints.
The EDM of the paramagnetic systems ($^{205}$Tl atom, YbF, and ThO molecules) are sensitive to the type 2 RPV bilinears, and also to the type 1 leptonic RPV bilinears ($x_1$ and $x_2$).
Thus, to be consistent with the constraint provided by the paramagnetic systems, it suffices to cancel the type 1 and type 2 contributions mutually.
As we can see, this is exactly what is happening  in Table \ref{table:rpvlpresult}.

The constraints from the EDM of diamagnetic atoms ($^{199}$Hg and $^{129}$Xe) also have strong sensitivity to type 2 RPV bilinears, since it is generated by P, CP-odd e-N interactions.
Diamagnetic atoms have, however, a moderate sensitivity to type 1 bilinears.
They also receive contribution from type 4 ($x_8 , x_9$ and $x_{10}$).
To be consistent with experimental upper bounds, we have to cancel the type 2 and type 4 contributions.
The large cancellation occurs among $x_3 ,x_4$ and $x_5$.
The remaining small part is cancelled with the type 4 RPV components originating in the nuclear Schiff moment.
Within the above constraints, it is possible to enlarge the $x_3$ (=$\sum {\rm Im} (\lambda_{i11}\lambda'^*_{i11})$) component up to $\sim 10^{-4}$.
The coefficients $c_{ia}$ of the EDM of diamagnetic atoms for type 2 RPV bilinears are aligned.
This explains the relatively small maximal prediction for the EDM of $^{211}$Rn atom.
The same analysis does not hold for the $^{225}$Ra atom, since the EDM of $^{225}$Ra has strong sensitivity on the hadronic sector (type 4).
The above inspection means that even by introducing the future experimental constraints from $^{211}$Rn and $^{225}$Ra atoms, which will be explained in Sec. \ref{subsec:prospective}, it is not possible to give tight upper bounds on RPV bilinears of type 2.
To rule out the type 2 bilinears, we need another observable with coefficients $c_{a3} , c_{a4}$ and $c_{a5}$ not aligned with those of the diamagnetic atoms.
This is possible  when we use the $R$-correlation of the neutron beta decay, which will be discussed later.

\subsubsection{Limits to type 3}

Throughout this analysis, the type 3 RPV bilinears are supposed to take the same value ($|x_6 | = 5.0 \times10^{-2}$, $|x_7 | = 5.5 \times10^{-2}$).
This is simply because the type 3 cannot be constrained from the linear programming analysis due to the suppression of the quark EDM by the electromagnetic coupling constant in the Barr-Zee type contribution.
The limits coming from other experiments are therefore dominant in this case.

\subsubsection{Hadronic observables}

The purely hadronic observables (such as EDMs of the neutron, proton, deuteron and $^3$He nucleus) have a large sensitivity against hadronic P, CP violating RPV interactions (type 4).
This is due to the absence of the screening electrons and also to the large sensitivity of the near-future experiments with novel techniques using the storage ring \cite{storage}.
The absence of the electrons also suppresses the semi-leptonic contribution, so that it serves to probe a fixed region of the RPV parameter space.
One of the important characteristics of the purely hadronic EDMs is that they depend approximately only on type 4 RPV bilinears (type 3 contribution is relatively small).
As they have restricted sensitivity against RPV bilinears, they can be used as a good probe to rule out a specific region of the RPV parameter space.
This also means that accumulating the EDM experimental data of pure hadronic EDMs is an efficient way to constrain the RPV parameter space, since they receive no cancellation from other sector than type 4 RPV bilinears.
We must note that the prediction of the hadronic EDMs suffers from large theoretical uncertainty due to the use of model calculations at the hadronic level, and we would not be quite sure even of the order of magnitude.
To do a quantitative analysis, we must improve the accuracy of the QCD level calculation.

\subsubsection{$R$-correlation}

With the assumption of the single coupling dominance, 
one might perhaps argue that the $R$-correlation is not particularly useful due to the strong constraint of the atomic EDMs against CP violation.
For example, within the dominance of one single bilinear of RPV couplings, the EDM of $^{199}$Hg atom can constrain the same combination Im$(\lambda_{i11} \lambda'^*_{i11})$ up to $10^{-10}$ with the current experimental data, whereas the $R$-correlation can constrain only up to $10^{-2}$ in the current experimental prospect of the $^8$Li \cite{murata}, well below the EDM experimental sensitivity.
The result of our analysis, however, shows the potential importance of this observable.
By virtue of the fact that the $R$-correlation depends only on one combination (at least at the tree level), it can be ``safely" large without conflicting with EDM observables which depend on several couplings.
The $R$-correlation is an important probe of the absolute size of Im($\lambda_{i11}\lambda'^*_{i11}$).
We have seen that Im($\lambda_{i11}\lambda'^*_{i11}$) is the most sensitive RPV bilinear for the atomic EDMs, and combined with the EDM-constraints of paramagnetic and diamagnetic atoms, we can fully constrain the type 2 RPV bilinears.
If the $R$-correlation can be measured with sufficient accuracy, it is possible to rule out the large prediction of atomic EDMs from Im($\lambda_{i11}\lambda'^*_{i11}$), thus reducing a large portion of the contribution to them.
The experimental development in searching for the $R$-correlation should therefore be strongly urged.

\subsubsection{Muon EDM}

The muon EDM is out of the scope of the present paper, but this observable sits in a special position, so we should add some comment.
The muon EDM depends on Im$(\lambda_{122} \lambda^*_{133})$, Im$(\lambda_{i22} \lambda'^*_{i22})$, and Im$(\lambda_{i22} \lambda'^*_{i33})$ via the two-loop level Barr-Zee type diagram, but these combinations do not contribute to the other available P, CP-odd observables.
In RPV, the muon EDM is actually completely independent of other observables and can constrain its own RPV parameter space exclusively.
As we do not have any other resource of EDM experimental data which can constrain the RPV couplings under consideration, the maximal prediction is just the sum of the upper bounds of RPV bilinears which can be probed with the muon EDM, and is of order $10^{-24} e\,{\rm cm}$.
The present experimental sensitivity ($\sim 10^{-19}e\, {\rm cm}$ \cite{muong2}) is well below the existing limits to the RPV couplings from other experiments.
The future muon EDM experiments is prepared to aim at the order of $10^{-24}e\, {\rm cm}$ \cite{storage}, but the maximal value predicted in this analysis is also of the same order.
Thus it will be difficult to either probe or constrain these RPV bilinears.

\subsubsection{Theoretical uncertainties}

Let us briefly mention theoretical uncertainties of this analysis.
The first source of large theoretical uncertainty is the nuclear level calculations.
The actual results of the Schiff moment calculations are not always consistent with each other (see Table \ref{table:schiffmoment}).
We have tested the dependence of the linear programming analysis on the different results presented in Ref. \cite{ban}.
The results may vary by one or two orders of magnitude.
This illustrates the large uncertainty due to the nuclear level calculation of the Schiff moments of nuclei with deformation.
The reduction of the theoretical uncertainty associated with odd numbered nuclei with deformation is one of the problems that challenges us.

The second large theoretical uncertainty comes from the hadron level calculation.
To derive the hadronic contribution of the chromo-EDM operator, we have used the chiral techniques, which have a strong dependence on the cutoff scale \cite{hisano,fuyuto} and are not considered to be accurate better than 100{\%}.
In the hadron sector, the final result may also change by an order of magnitude.
To reduce these theoretical uncertainties, the lattice QCD study is absolutely needed \cite{bhattacharya}.

By considering these sources of theoretical error, we have to say that coefficients $c_6$, $c_7$, $c_8$, $c_9$ and $c_{10}$ could differ by an order of magnitude.

\subsubsection{Improvement of constraints on RPV couplings from other experiments}

We must note that the limits on RPV couplings can be tightened possibly by improving the constraints provided by other experiments in the linear programming analysis.
This is because the reductions of the allowed region of the (absolute values of) RPV couplings can be expected to constrain the degrees of freedom left for the rearrangement of parameters within (EDM-)constraints.
Here we mention the possible potentiality of other experiments.

The first possibility is to improve the experimental accuracy of the measurements of lepton decays (universality test) and hadron decays ($K\rightarrow \pi \nu \bar \nu$, $B\rightarrow X_s \nu \bar \nu $) (see Table \ref{table:rpvlimit}).

The second possibility is the constraint from the absolute mass of neutrinos.
All lepton number violating RPV interactions contribute to the Majorana mass of the neutrinos, so the improvement of experimental constraints on their masses has a large potential to limit RPV bilinears relevant in our analysis.

The third interesting possibility is the limit to be afforded by collider experiments.
Actually, resonances of sneutrino arise in the presence of the RPV interactions at collider experiments \cite{rpvcollider}.
The parton distributions for strange and bottom quarks also allow to probe the RPV interactions $\lambda'_{i22}$ and $\lambda'_{i33}$.
The lepton collider is sensitive to the resonances of sneutrino generated by leptonic RPV interactions $\lambda_{i11}$ (electron collision), $\lambda_{i22}$ (muon collision) and $\lambda_{i33}$ ($\tau$ lepton collision).

\subsection{\label{subsec:prospective}Prospective upper limits of EDMs}

Finaly we would like to supplement a rather peripheral analyses. 
Namely we add an estimate of the constraints on RPV bilinears when the prospective upper limits of the EDMs of $^{129}$Xe,  neutron, proton, deuteron, $^3$He nucleus, $^{211}$Rn, $^{225}$Ra, and $^{210}$Fr atoms are imposed. We will set the following limits, regarding the prospects of the experimental sensitivity \cite{storage,bnl,mueller,sakemi,ucn,xeasahi}:
\begin{eqnarray}
|d_n | &<& 10^{-28} e \, {\rm cm} \ , \nonumber\\
|d_{\rm Xe} | &<& 10^{-31} e \, {\rm cm} \ , \nonumber\\
|d_p | &<& 10^{-29} e \, {\rm cm} \ , \nonumber\\
|d_d | &<& 10^{-29} e \, {\rm cm} \ , \nonumber\\
|d_{\rm He} | &<& 10^{-29} e \, {\rm cm} \ , \nonumber\\
|d_{\rm Rn} | &<& 10^{-29} e \, {\rm cm} \ , \nonumber\\
|d_{\rm Ra}| &<& 3 \times 10^{-28} e \, {\rm cm} \ , \nonumber \\
|d_{\rm Fr} | &<& 1\times10^{-26} e \, {\rm cm} .
\label{eq:additional}
\end{eqnarray}

Let us first see the limits on RPV bilinears with a combined use of (\ref{eq:additional}) and the linear programming method  for the six EDM-constraints ($^{205}$Tl, $^{199}$Hg, $^{129}$Xe, YbF, ThO and neutron) seen previously.
The result is shown in Table \ref{table:rpvlpresultplusalpha}.

\begin{table*}
\begin{center}
\begin{tabular}{c|ccccccccc}
\hline \hline
RPV bilinear &$d_n$ &$d_{\rm Xe}$ & $d_p$&$d_d$&$d_{\rm He}$&$d_{\rm Rn}$&$d_{\rm Ra}$&$d_{\rm Fr}$& All \\
\hline
$|x_1|$& 0.15 &0.15&0.15& 0.15&0.15&0.15&0.15&0.15&0.15\\
$|x_2|$& 0.12 &$6.9\times 10^{-2}$ &0.12&$6.6\times 10^{-2}$&$6.6\times 10^{-2}$&$6.1\times 10^{-2}$&$6.5\times 10^{-2}$&$3.2\times 10^{-2}$&$1.3\times 10^{-2}$\\
$|x_3|$&$1.8\times 10^{-4}$ &$1.0\times 10^{-4}$&$1.8\times 10^{-4}$&$8.1\times 10^{-5}$&$8.1\times 10^{-5}$&$7.2\times 10^{-5}$&$8.1\times 10^{-5}$&{\colourcyan{$1.9\times 10^{-5}$}}&$1.2\times 10^{-7}$\\
$|x_4|$&$1.1\times 10^{-2}$&$5.4\times 10^{-3}$&$1.1\times 10^{-2}$&$6.1\times 10^{-3}$&$6.1\times 10^{-3}$&$5.6\times 10^{-3}$&$6.0\times 10^{-3}$&$2.7\times 10^{-3}$&$6.1\times 10^{-6}$\\
$|x_5|$& 0.19 &0.10&0.19 &$9.5\times 10^{-2}$&$9.5\times 10^{-2}$& $8.6\times 10^{-2}$ &$9.5\times 10^{-2}$&$3.3\times 10^{-2}$&$7.0\times 10^{-5}$\\
$|x_6|$&$5.0\times 10^{-2}$&$5.0\times 10^{-2}$&$5.0\times 10^{-2}$&$5.0\times 10^{-2}$&$5.0\times 10^{-2}$&$5.0\times 10^{-2}$&$5.0\times 10^{-2}$&$5.0\times 10^{-2}$&$5.0\times 10^{-2}$\\
$|x_7|$&$5.5\times 10^{-2}$&$5.5\times 10^{-2}$&$5.5\times 10^{-2}$&$5.5\times 10^{-2}$&$5.5\times 10^{-2}$&$5.5\times 10^{-2}$&$5.5\times 10^{-2}$&$5.5\times 10^{-2}$&$4.2\times 10^{-3}$\\
$|x_8|$&$8.9\times 10^{-3}$&$8.9\times 10^{-3}$&$8.9\times 10^{-3}$&$8.9\times 10^{-3}$&$8.9\times 10^{-3}$&$8.9\times 10^{-3}$&$8.9\times 10^{-3}$&$8.9\times 10^{-3}$&$2.5\times 10^{-6}$\\
$|x_9|$&$3.0\times 10^{-2}$&$3.0\times 10^{-2}$&$3.0\times 10^{-2}$&$9.0\times 10^{-4}$&$9.6\times 10^{-4}$&$3.3\times 10^{-3}$&$9.7\times 10^{-4}$&$3.0\times 10^{-2}$&$2.2\times 10^{-7}$\\
$|x_{10}|$& 0.15 & 0.15 &0.15 &$3.1\times 10^{-3}$&$3.6\times 10^{-3}$&$1.5\times 10^{-2}$&$3.7\times 10^{-3}$&0.15 &$6.6\times 10^{-7}$\\
\hline
\end{tabular}
\end{center}
\caption{
Upper limits on the absolute value of bilinears of RPV couplings found by linear programming analysis within the EDM-constraints of neutron, $^{205}$Tl, $^{199}$Hg, $^{129}$Xe atoms, YbF, ThO molecules plus one additional limit from prospective EDM experiment of Eq. (\ref{eq:additional}).
In the final row, we have given the upper limits for the RPV bilinears when the all prospective constraints of Eq. (\ref{eq:additional}) were applied simultaneously.
The sparticle mass was set to 1 TeV.
}
\label{table:rpvlpresultplusalpha}
\end{table*}

By comparing Tables \ref{table:rpvlpresult1} and \ref{table:rpvlpresultplusalpha}, we see that many RPV bilinears ($x_2$, $x_3$, $x_4$, $x_5$, $x_7$, $x_8$, $x_9$ and $x_{10}$) can 
further be constrained by considering experimental data  that are expected to come in the future.
This result shows the importance of the prospective experiments.
The following observations can be done:

\begin{itemize}

\item
RPV bilinears $x_1$ and $x_6$ cannot be constrained at all.
This is due to the small contribution of the Barr-Zee type diagram with the second generation fermion in the inner loop (the Barr-Zee type diagram is sensitive 
to the inner loop fermion mass).

\item
The  expected bound on the proton EDM in (\ref{eq:additional})
is only effective for limiting $x_7$, in spite of the strong dependence of the proton EDM on the purely hadronic bilinears (type 4). This is due to the alignment of the coefficients with the neutron EDM.

\item
By taking into account of the EDM-constraints of the deuteron, 
$^3$He nucleus and $^{225}$Ra atom in (\ref{eq:additional}), the hadronic RPV bilinears $x_9$ and $x_{10}$ are constrained by two orders of magnitude tighter.
This shows the strong sensitivity of hadronic EDMs (deuteron and $^3$He) against $x_9$ and $x_{10}$.
The EDM of $^{225}$Ra atom is also sensitive to the hadronic $R$-parity violation, due to the strong enhancement of the nuclear Schiff moment.

\item
The leptonic and semi-leptonic RPV bilinears $x_2$, $x_3$ and $x_4$, although moderate, can be constrained with any additional prospective EDM constraints.
Na\"{i}vely, this fact is not obvious, since the purely hadronic EDMs ($d_d$ and $d_{\rm He}$) are not sensitive to  $x_2$, $x_3$ and $x_4$.
This can be understood by the interplay between the additional future EDM-constraints and the existing constraints of diamagnetic atoms ($^{199}$Hg and $^{129}$Xe).
This fact shows the importance of considering as many systems of
the experimental EDM-constraints as possible.

\item
The experimental limit of the $^{211}$Rn EDM can give tighter constraints on $x_2$, $x_3$, $x_4$, $x_9$ and $x_{10}$.
This is realized thanks to the strong (prospective) EDM-constraints.

\item
As we see from Tables \ref{table:rpvlpresult1} and \ref{table:rpvlpresultplusalpha}, 
improved measurements of neutron EDM $d_{n}$ will not necessarily provide us 
with more precise information of $x_{i}$'s.  This is simply explained 
by the fact that the location of the maximum point in the parameter space 
is away from the hyperplane determined by the constraints of the neutron 
dipole moment. 

\end{itemize}

We have also given upper limits when all prospective EDM-constraints are applied simultaneously, and the result gives very strong limits on RPV couplings.
We understand that, when the number of constraints exceeds the number 
of the RPV bilinears, these strong limitations often occur.
It is to be noted that by considering all prospective limits, we can constrain the RPV bilinear $x_8$.
We can conclude from these results that the combination of many EDM-constraints are useful in setting upper bounds on the RPV interactions.

\section{\label{sec:conclusion}Conclusion}

In summary, we have done an extensive analysis having a careful look at every corner of the full CP violating RPV parameter space.
We have developed a new calculational technique based on the linear programming method, have given limits on the imaginary parts of RPV bilinears, and have predicted the maximally possible values for observables of prepared or on-going experiments (proton, deuteron, $^3$He nucleus, $^{211}$Rn, $^{225}$Ra, $^{210}$Fr atoms, muon, and the $R$-correlation of the neutron beta decay) under the currently available experimental constraints (\ref{eq:currentexpupperlimit}) (in addition to other CP conserving experimental data of fundamental precision tests listed in Table \ref{table:rpvlimit}).
We have found through this analysis that the RPV bilinears $x_2[= {\rm Im}(\lambda_{211}\lambda^*_{233}) ]$, $x_3[= \sum_{i=2,3} {\rm Im}(\lambda_{i11}\lambda'^*_{i11}) ]$, $x_4[= \sum_{i=2,3} {\rm Im}(\lambda_{i11}\lambda'^*_{i22}) ]$, $x_9[= \sum_{i=1,2,3} {\rm Im}(\lambda'_{i11}\lambda'^*_{i33}) ]$ and $x_{10}[= \sum_{i=1,2,3} {\rm Im}(\lambda'_{i22}\lambda'^*_{i33}) ]$ can be constrained by the current available experimental EDM-constraints.

The upper limits for $x_{i}$'s obtained by linear programming method 
differ from those obtained under the hypothesis of 
the single-coupling dominance by a few orders of magnitudes.
The larger upper limits in the linear programing method has been made possible 
 because of the destructive interference among the terms in Eq.
(\ref{eq:sum}),  thereby 
 we have looked for the maximum parameter space of $x_i$
within the constrains of EDM data.
On the other hand, destructive cancellation 
in Eq.  (\ref{eq:sum})  was avoided 
as much as possible when we looked for maximally possible values of EDMs 
of proton, deuteron, He, Rn, Ra, and Fr 
in Table \ref{table:rpvlpresult}
and Table \ref{table:rpvlpresult5TeV}.
Only for the EDM of proton and Fr, the cancellation is  of two and one digits respectively.

When we look at the numbers of $x_{i}$ in Table \ref{table:rpvlpresult}, 
we notice that $x_{3}$ is much smaller than other $x_{i}$'s. This means that 
among the two terms in the sum $x_{3}=\sum _{i=2, 3}{\rm Im}(\lambda _{i11} \lambda_{i11}^{\prime *} )$ large cancellation could admittedly  be occurring.

The upper limits on the above-mentioned  RPV-bilinears 
$x_{2}$, $x_{3}$, $x_{4}$, $x_{9}$, and $x_{10}$ together with 
$x_5[= \sum_{i=2,3} {\rm Im}(\lambda_{i11}\lambda'^*_{i33}) ]$, $x_7 [= \sum_{i=1,2} {\rm Im}(\lambda_{i33}\lambda'^*_{i11}) ]$, and $x_8 [= \sum_{i=1,2,3} {\rm Im}(\lambda'_{i11}\lambda'^*_{i22})]$ can be tightened with additional prospective EDM-constraints of the proton, deuteron, $^3$He nucleus, $^{211}$Rn and $^{225}$Ra atoms given in (\ref{eq:additional}).
In particular, $x_9$ and $x_{10}$ can be strongly constrained due to the high sensitivity of the planned EDM experiments.
RPV bilinears $x_1[= {\rm Im}(\lambda_{311}\lambda^*_{322})] $ and $x_6[= \sum_{i=1,3} {\rm Im}(\lambda_{i22}\lambda'^*_{i11}) ]$ have not yet been constrained due to the rather poor sensitivity of the EDMs of the relevant systems.

For the prediction of prospective experiments, we have found that very large values are still allowed.
This result is encouraging for experimentalists, since there is still a possibility to observe large EDM for prospective experiments.
We have made a comparison of our analyses with the ``classic" ones which assume the dominance of only one or a few combinations of couplings among several others.
We have demonstrated the potential importance of the regions in the parameter space which had never been given due consideration.
As it has turned out in this work, intriguing observables are the $R$-correlation, the purely hadronic EDMs and the muon EDM which have sensitivity to the restricted area of the RPV parameter space.
We have also obtained the useful information that the $R$-correlation is an important probe to rule out the type 2 RPV bilinears.
Although we believe that applications of linear programming method in this work is quite successful, we have to admit some troubles due to the theoretical uncertainties.
The reduction of them, in particular at the hadronic and nuclear level, are urgently required.

Finally we would like to mention a subject which is left open 
for our future work.
In the present work, we have analyzed the EDM-constraints as linear relations in the linear programming, but the absolute limits on the RPV couplings taken from other experimental data were assumed to hold only for a single RPV coupling.
As a future subject, we have to treat also these absolute limits offered by other experiments as linear relation inputs in the analysis of linear programming.

\begin{acknowledgments}
The authors thank T. Hatsuda for useful discussion and comments.
This work is in part supported by the Grant for Scientific Research 
[Priority Areas ``New Hadrons'' (E01:21105006), (C) No.23540306] from the Ministry of Education, Culture, Science and Technology 
(MEXT) of Japan and JSPS KAKENHI Grant No. 24540273 and 25105010(T.S.).
This is also supported by the RIKEN iTHES Project.
\end{acknowledgments}

\appendix

\section{\label{sec:barr-zee}Barr-Zee type contribution to the fermion EDM and quark chromo-EDM}

The contribution to the EDM of a fermion $F_{k}$ due to the photon exchange RPV Barr-Zee type graph (see Fig. \ref{fig:barr-zee}) is given by \cite{rpvedmsfermion}
\begin{eqnarray}
d_{\rm BZ}  (j,k,Q_f,Q_F)
&=&
d^\gamma_{\rm BZ}  (j,k,Q_f,Q_F) 
\nonumber\\
&&
+
d^Z_{\rm BZ}  (j,k,Q_f,Q_F)
\nonumber\\
&&
+
d^W_{\rm BZ}  (j,k,Q_f,Q_F),
\label{eq:Barr-Zee1}
\end{eqnarray}
where $k$ and $Q_F$ are respectively the flavor index and the electric charge in unit of $e$ of the external fermion $F_k$. 
Likewise $j$ and $Q_f$ are respectively the flavor index and electric charge in unit of $e$ of the the inner loop fermion $f_j$ (or sfermion $\tilde f_j$).
In the left-hand side of 
(\ref{eq:Barr-Zee1}), $d_{F_k}^\gamma$, $d_{F_k}^Z$, and $d_{F_k}^W$ are the Barr-Zee type contributions with photon, $Z$ and $W$ boson exchange, respectively, to be defined below.
Note that the electric charge $e=|e|$ is positively defined, in contrast to Refs. \cite{yamanakabook,yamanaka1,rpvedmsfermion}.

The photon exchange RPV effect is 
\begin{eqnarray}
d^\gamma_{\rm BZ}  (j,k,Q_f,Q_F) 
&\equiv &
\frac{ \alpha_{\rm em} n_c Q_f^2 Q_F e}{16\pi^3} \cdot \frac{m_{f_j}}{ m_{\tilde \nu_i }^2 }  
\nonumber\\
&& \times
\Biggl[
F \left( \frac{m_{f_j}^2 }{ m_{\tilde \nu_i}^2 } \right)
-\frac{1}{2} F \left( \frac{m_{\tilde f_{Lj}}^2 }{ m_{\tilde \nu_i}^2 } \right)
\nonumber\\
&& \hspace{5em}
-\frac{1}{2} F \left( \frac{m_{\tilde f_{Rj}}^2 }{ m_{\tilde \nu_i}^2 } \right)
\Biggr]
 , \ \ \ \ 
\label{eq:Barr-Zee2}
\end{eqnarray}
where $n_c$ is the color number of the inner loop fermion or sfermion ($n_c=3$ if inner loop fermion is a quark, otherwise $n_c=1$).
The mass of the exchanged sneutrino is given by $m_{\tilde \nu_i}$.
The function $F$ is defined as
\begin{eqnarray}
F(\tau ) 
&=&
 \int^1_0 dx \frac{x(1-x)}{ x(1-x) - \tau} \ln \left( \frac{ x(1-x)}{\tau} \right) 
\nonumber\\
&\approx& 
-2-\ln \tau 
.
\end{eqnarray}
The last approximation holds for small $\tau$.
Note that the Barr-Zee type diagram gives EDM contribution only to down-type quarks (the same property holds also for the chromo-EDM seen below).

The contribution to the EDM of the fermion $F_k$ from the $Z$ boson exchange RPV Barr-Zee type graph is given by \cite{rpvedmsfermion}
\begin{eqnarray}
d^Z_{\rm BZ}  (j,k,Q_f,Q_F)
&\equiv &
\frac{n_c Q_f \alpha_F e \alpha_{\rm em}}{32\pi^3 } m_{f_j} 
\nonumber\\
&&
\times \int_0^1 \hspace{-.5em} dz \Biggl\{
2\alpha_f  I \Bigl(m_Z^2 \, , \,  m_{\tilde \nu_i}^2 \, , \frac{m_{f_j}^2}{z(1-z)} \Bigr)
\nonumber\\
&&
-\hspace{-1.2em}\sum_{\tilde f_j = \tilde f_{Lj},\tilde f_{Rj}} \hspace{-1em} \alpha_{\tilde f}  I \Bigl(m_Z^2 \, , \,  m_{\tilde \nu_i}^2 \, , \frac{m_{\tilde f_j}^2}{z(1-z)} \Bigr)
\Biggr\}
, 
\nonumber\\
\label{eq:mzbz}
\end{eqnarray}
where the weak coupling $\alpha_f \equiv \frac{1}{4} (3\tan \theta_W -\cot \theta_W) \approx -0.065$ for $f$ being a charged lepton, and $\alpha_f \equiv \frac{1}{12} \tan \theta_W -\frac{1}{4}\cot \theta_W \approx -0.42$ for $f$ being a  down-type quarks.
For the sfermion weak coupling, we have $\alpha_{\tilde f_L}= \alpha_f -\beta_f$ and $\alpha_{\tilde f_R}= \alpha_f +\beta_f$, where $\beta_f \equiv \frac{1}{4}(\tan \theta_W + \cot \theta_W)$ for $f$ being a charged lepton or down-type quark.
The integral $I(a,b,c)$ is defined as
\begin{eqnarray}
I(a,b,c) 
&=&
\int _{0}^{\infty}\frac{xdx}{(x+a)(x+b)(x+c)}
\nonumber \\
&=&
\frac{1}{(b-a)(c-b)(a-c)}\nonumber\\
&& \times 
 \left[ \ ab \ln \left| \frac{a}{b} \right| + bc \ln \left| \frac{b}{c} \right| +ca \ln \left| \frac{c}{a} \right| \ \right]
 . \ \ \ \ 
\label{eq:integformula}
\end{eqnarray}

The $W$ exchange RPV Barr-Zee diagram contribution turns out to be \cite{rpvedmsfermion}
\begin{eqnarray}
d^W_{\rm BZ}  (j,k,Q_f,Q_F)
&\equiv &
\frac{n_ce \alpha_{\rm em} V_{jj} V_{kk} m_{f^d_j}}{128\pi^3 \sin^2 \theta_W} 
\nonumber\\
&& \times 
\int_0^1 \hspace{-.5em}dz \, (Q_u (1-z)+Q_d z)
\nonumber\\
&&
\times
\Biggl[
I \Bigl(m_W^2 \, , \,  m_{\tilde e_{Li}}^2 \, , \, \frac{ m_{f^u_j}^2}{z} +\frac{m_{f^d_j}^2}{1-z} \Bigr)  
\nonumber\\
&& 
-I \Bigl(m_W^2 \, , \,  m_{\tilde e_{Li}}^2 \, , \, \frac{ m_{\tilde f^u_j}^2}{z} +\frac{m_{\tilde f^d_j}^2}{1-z} \Bigr)  
\Biggr]
,
\nonumber\\
\label{eq:dwinteg}
\end{eqnarray}
where $V_{ij}$ is the Cabibbo-Kobayashi-Maskawa matrix element, $f^d_j$ and $f^u_j$ denote the down- and up-type inner loop fermions, respectively.
The mass of the selectron is denoted by $m_{\tilde e_{Li}}$.
Here we have again used the integral of Eq. (\ref{eq:integformula}).

The Barr-Zee type graph contribution to the down-type quark chromo-EDM is 
\begin{eqnarray}
d_{\rm BZ}^c (j,k)
&\equiv &
 \frac{ \alpha_s }{32\pi^3} \cdot \frac{m_{q_j}}{ m_{\tilde \nu_i }^2 } 
\nonumber\\
&& \times
\Biggl[
F \left(  \frac{m_{q_j}^2 }{ m_{\tilde \nu_i}^2} \right) 
-\frac{1}{2} F \left(  \frac{m_{\tilde q_{Lj}}^2 }{ m_{\tilde \nu_i}^2} \right) 
\nonumber\\
&& \hspace{6em}
-\frac{1}{2} F \left(  \frac{m_{\tilde q_{Rj}}^2 }{ m_{\tilde \nu_i}^2} \right) 
\Biggr],
\label{eq:chromo-Barr-Zee1}
\end{eqnarray}
where $k$ and $j$ are again the flavor indices of the quark $q_k$ and the fermion of the inner loop, respectively.

\section{\label{sec:pseudoscalar}Phenomenological pseudoscalar contents of nucleon}

The derivation of the pseudoscalar condensates goes along the line of Ref. \cite{cheng}.
The anomalous Ward identity requires
\begin{eqnarray}
2m_N \Delta u &=& 2m_u \langle p| \bar ui\gamma_5 u |p\rangle - 2 m_N \frac{\alpha_s }{2 \pi} \Delta g ,
\\
2m_N \Delta d &=& 2m_d \langle p| \bar di\gamma_5 d |p\rangle - 2 m_N \frac{\alpha_s }{2 \pi} \Delta g ,
\\
2m_N \Delta s &=& 2m_s \langle p| \bar si\gamma_5 s |p\rangle - 2 m_N \frac{\alpha_s }{2 \pi} \Delta g ,
\end{eqnarray}
where $\Delta u = 0.71 \pm 0.04$, $\Delta d = -0.38 \pm 0.06$, and $\Delta s = -0.02 \pm 0.01$ \cite{compass}.
The anomaly contribution is defined as $\Delta g = -\frac{1}{4m_N} \langle N | G_{\mu \nu}^a \tilde G^{\mu \nu}_a | N \rangle$, and is given by
\begin{eqnarray}
\frac{\alpha_s }{2 \pi} \Delta g 
&=&
\left( \frac{2/3}{1+F/D} -\frac{1}{1+m_u /m_d} \right) g_A
\nonumber\\
&&
+\frac{m_u / m_d}{m_u^2 /m_d^2 -1} \delta g_A 
-\frac{1}{3} \Delta \Sigma
\nonumber\\
&\approx &
-0.30
,
\label{eq:gluon_spin}
\end{eqnarray}
where $\Delta \Sigma = 0.32 \pm 0.03 \pm 0.03$ \cite{compass}, $D=0.80$ and $F=0.47$.
Here $\delta g_A = g_A - \frac{f_\pi}{m_N} g_{\pi NN}$ is the correction to the Goldberger-Treiman relation, where we have used $g_{\pi NN} = 14.11 \pm 0.20$ \cite{ericson}.
By equating the above equations, we obtain Eqs. (\ref{eq:u5u}), (\ref{eq:d5d}), and (\ref{eq:s5s}).

\section{\label{sec:heavy_quark}Heavy quark contents of nucleon}

Here we calculate the heavy quark content of the nucleon.
The scalar content of nucleon satisfies the following relation
\begin{equation}
m_N
=
\frac{\beta_{\rm QCD}}{2 g_s} \langle N | G_{\mu \nu}^a G^{\mu \nu}_a | N \rangle
+ \sum_q m_q \langle N | \bar qq | N \rangle
,
\label{eq:trace_anomaly}
\end{equation}
where $\beta_{\rm QCD} = - \beta_0 \frac{g_s^3 }{16 \pi^2} + O (g_s^5)$ with $\beta_0 =9$.
The first term on the right hand side of (\ref{eq:trace_anomaly}) is the contribution in the chiral limit, and the other terms the finite quark mass contribution to the nucleon mass.
In the heavy quark expansion, the scalar heavy quark condensates are given by \cite{zhitnitsky}
\begin{equation}
\langle N | \bar QQ | N \rangle
=
-\frac{\alpha_s }{12\pi m_Q } \langle N | G_{\mu \nu}^a G^{\mu \nu}_a | N \rangle
+ O \left( \frac{1}{m_Q^2} \right)
,
\label{eq:scalar_HQ_expansion}
\end{equation}
where $Q = c,b,t$.
By equating Eqs. (\ref{eq:trace_anomaly}) and (\ref{eq:scalar_HQ_expansion}), we obtain (renormalization scale $\mu = 2$ GeV)
\begin{eqnarray}
\frac{\beta_{\rm QCD}}{2 g_s} \langle N | G_{\mu \nu}^a G^{\mu \nu}_a | N \rangle
&\approx &
690 {\rm MeV}
,
\\
\langle N | \bar cc | N \rangle
&\approx &
4.0 \times 10^{-2}
,
\\
\langle N | \bar bb | N \rangle
&\approx &
1.2 \times 10^{-2}
,
\\
\langle N | \bar tt | N \rangle
&\approx &
2.9 \times 10^{-4}
,
\end{eqnarray}
where we have used the result of the lattice QCD study of the nucleon sigma term $m_u \langle N | \bar uu | N \rangle + m_d \langle N | \bar dd | N \rangle \approx 2 \times 45$ MeV \cite{lattice}, and numerical values (\ref{eq:uu}) - (\ref{eq:ss}) input parameters.
The result of the charm quark content $\langle N | \bar cc | N \rangle$ is to be compared with the recent lattice QCD result $\langle N | \bar cc | N \rangle \approx 0.03 - 0.09$ \cite{lattice_charm_content}, which is on the same order of magnitude.

The pseudoscalar heavy quark contents of nucleon can be written in the leading order of the heavy quark expansion \cite{zhitnitsky} as
\begin{eqnarray}
\langle N | \bar Q i \gamma_5 Q | N \rangle
&=&
-\frac{\alpha_s}{8 \pi m_Q}
\langle N | G_{\mu \nu}^a \tilde G^{\mu \nu}_a | N \rangle
+ O\left( \frac{1}{m_Q^2} \right)
\nonumber\\
&=&
\frac{m_N}{m_Q} \frac{\alpha_s}{2\pi} \Delta g
+ O\left( \frac{1}{m_Q^2} \right)
.
\end{eqnarray}
By using Eq. (\ref{eq:gluon_spin}), we obtain 
\begin{eqnarray}
\langle N | \bar ci\gamma_5 c | N \rangle
&\approx &
-0.22
,
\\
\langle N | \bar bi\gamma_5 b | N \rangle
&\approx &
-6.8 \times 10^{-2}
,
\\
\langle N | \bar ti\gamma_5 t | N \rangle
&\approx &
-1.7 \times 10^{-3}
.
\end{eqnarray}

\section{\label{sec:appendixnucleonedm}Quark chromo-EDM contribution to the nucleon EDM}

The leading contribution of the quark chromo-EDM to the neutron and proton EDMs is given by the chiral logarithm of the meson-loop diagrams shown in Figs. \ref{fig:neutronmesonloop} and \ref{fig:protonmesonloop}.
The dependence of the nucleon EDMs on the quark chromo-EDM is given by \cite{hisano,fuyuto,yamanakabook}
\begin{eqnarray}
d_n
&=& 
-\frac{e}{4\sqrt{2} \pi^2 f_\pi } 
\nonumber\\
&& \times
\left[ 
(D+F) \bar g_{pn \pi} \ln \frac{\Lambda}{m_\pi} 
-(D-F) \bar g_{\Sigma n K} \ln \frac{\Lambda}{m_K} 
\right]
 , \nonumber\\
d_p
&=&
\frac{e}{4\sqrt{2} \pi^2 f_\pi } 
\nonumber\\
&& \times
\Biggl[ 
(D+F) \bar g_{pn \pi} \ln \frac{\Lambda}{m_\pi} 
+\frac{D-F}{\sqrt{2}} \bar g_{\Sigma p K} \ln \frac{\Lambda}{m_K} 
\nonumber\\
&& \hspace{2em}
+\frac{D+3F}{\sqrt{6}} \bar g_{\Lambda n K} \ln \frac{\Lambda}{m_K} 
\Biggr]
,
\end{eqnarray}
with
\begin{eqnarray}
\bar g_{pn \pi} 
&=&
-\frac{1}{2\sqrt{2} f_\pi } 
\Bigl[
D_u -D_d -m_0^2 (S_u -S_d)
\Bigr]
(d^c_u+d^c_d)
,
\nonumber\\
\bar g_{ \Sigma n K} 
&=&
-\frac{1}{2\sqrt{2} f_\pi } 
\Bigl[
D_s -D_d -m_0^2 (S_s -S_d)
\Bigr]
(d^c_u+d^c_s)
,
\nonumber\\
\bar g_{ \Sigma p K} 
&=&
\frac{\bar g_{ \Sigma n K}  }{\sqrt{2}}
,
\nonumber\\
\bar g_{ \Sigma n K} 
&=&
-\frac{1}{2\sqrt{2} f_\pi } (d^c_u+d^c_s )
\nonumber\\
&& \times
\frac{1}{\sqrt{6}}
\Bigl[
D_d +D_s -2 D_u -m_0^2 (S_d + S_s -2 S_u)
\Bigr]
,
\nonumber\\
\end{eqnarray}
where  $D=0.80$, $F=0.47$, $S_q \equiv \langle p | \bar q q | p \rangle $, $D_q \equiv \langle p | g_s \bar q \sigma_{\mu \nu} t_a G_a^{\mu \nu} q | p \rangle $, and $m_0^2 \equiv \frac{\langle 0 | g_s \bar q \sigma_{\mu \nu} t_a G_a^{\mu \nu} q | 0 \rangle}{\langle 0 | \bar q q | 0 \rangle}$.
By assuming $\langle p | g_s \bar q \sigma_{\mu \nu} t_a G_a^{\mu \nu} q | p \rangle \propto \langle p | \bar q q | p \rangle$, we find Eq. (\ref{eq:rhosp_rhosn}).

\section{\label{sec:r-corr}RPV contribution to the $R$-correlation}

We briefly present the $R$-correlation of the neutron beta decay, a powerful probe of the new physics with the small SM background \cite{betadecaybeyondsm,ckmbetadecay}.
This observable, like the EDMs, is also sensitive to the P and CP violations of the underlying theory.
The angular distribution of the neutron beta decay can be written as \cite{jacksontreiman}
\begin{eqnarray}
&&\omega (E_e , \Omega_e , \Omega_{\nu})
\propto 
1 + a \frac{{\vec p_e} \cdot {\vec p_{\nu}}}{E_e E_{\nu}}
+b \frac{m_e}{E_e} \nonumber \\
&&+ \ \vec \sigma_n \ \cdot \Biggl[
A\frac{{\vec p_e}}{E_e} + B\frac{{\vec p_{\nu}}}{E_{\nu}}
+D\frac{{\vec p_e}\times {\vec p_{\nu}}}{E_e E_{\nu}} \Biggr] \nonumber\\
&&+\ \vec \sigma_e \ \cdot \Biggl[ N \vec \sigma_n
+Q \frac{\vec p_e}{E_e +m_e} \frac{ \vec \sigma_n \cdot \vec p_e}{E_e}
+ R \frac{ \vec \sigma_n \times \vec p_e}{E_e} \Biggr]  + \cdots\ , \nonumber\\
\label{eq:decay_distribution}
\end{eqnarray}
where $R$-correlation is the last term, the triple product of the initial neutron polarization, emitted electron polarization and momentum.
This observable is odd under P and CP.

\begin{figure}[h]
\begin{center}
\includegraphics[height=28mm]{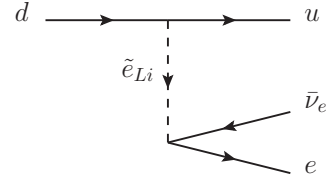}
\caption{\label{fig:rpvr-corr} 
$R$-parity violating contribution to the d-quark beta decay at the tree level.
}
\end{center}
\end{figure}

The tree level RPV contribution to the $R$-correlation of the neutron beta decay is given by the diagram given in Fig. \ref{fig:rpvr-corr}.

The nucleon level effective interaction is then
\begin{equation}
H_{\beta } \approx -\sum_{i=2,3} g_S \frac{\lambda_{i11} \lambda'^*_{i11}}{4m_{\tilde e_{Li}}^2} \bar pn \cdot \bar e (1-\gamma_5 ) \nu_e \ ,
\end{equation}
where $g_S \equiv \langle p| \bar u d |n\rangle \approx \langle p| \bar uu - \bar dd  |p\rangle \approx 1.02$ has been obtained by taking the isospin breaking to the first order \cite{alonso,rpvbetadecay}.
The lattice QCD study of $g_S$ also gives a consistent result \cite{bhattacharya2}.
We finally obtain the $R$-correlation formula presented previously in (\ref{eq:Rc3x3}) and (\ref{eq:rcoef}) 
\begin{equation}
R = \frac{g_A \langle p| \bar uu - \bar dd  |p\rangle}{2V_{ud} (1+ 3g_A^2) \frac{G_F}{\sqrt{2}} m_{\tilde e_{Li}}^2 } \sum_{i=2,3} {\rm Im } (\lambda_{i11} \lambda'^*_{i11}) \ , 
\end{equation}
where $g_{A}$ is given by (\ref{eq:gagvratio}).
One of the important characteristics of the $R$-correlation relevant in this discussion is that it depends only on the RPV combinations $\sum_{i=2,3}$Im($\lambda_{i11} \lambda'^*_{i11}$).

\end{document}